\documentclass[]{aa}  

\usepackage{graphicx}
\usepackage{txfonts}
\usepackage[]{hyperref}
\usepackage{natbib} 
\bibpunct{(}{)}{;}{a}{}{,} 

\usepackage{verbatim}
\usepackage{url}
\usepackage{epstopdf}

\begin{document}

   \title{Overshooting calibration and age determination from evolved binary systems} 

   \subtitle{A statistical investigation on biases and random variability}

   \author{G. Valle \inst{1,2,3}, M. Dell'Omodarme \inst{3}, P.G. Prada Moroni
     \inst{2,3}, S. Degl'Innocenti \inst{2,3} 
          }
   \titlerunning{Calibration of free parameters from binary stars}
   \authorrunning{Valle, G. et al.}

   \institute{
INAF - Osservatorio Astronomico di Collurania, Via Maggini, I-64100, Teramo, Italy 
\and
 INFN,
 Sezione di Pisa, Largo Pontecorvo 3, I-56127, Pisa, Italy
\and
Dipartimento di Fisica "Enrico Fermi'',
Universit\`a di Pisa, Largo Pontecorvo 3, I-56127, Pisa, Italy
 }

   \offprints{G. Valle, valle@df.unipi.it}

   \date{Received //; accepted //}

  \abstract
 {}
{ The capability of grid-based techniques to estimate the age together with the convective core overshooting efficiency of stars in
        detached eclipsing binary systems for main sequence stars has previously been investigated. We have extended this investigation to later evolutionary stages and have evaluated the bias and variability on the recovered age and convective core overshooting parameter accounting for both observational and internal uncertainties.}
{
 We considered synthetic binary systems, whose age and overshooting efficiency should be recovered by applying the SCEPtER pipeline to the same grid of models used to build the mock stars.
 We focus our attention on a binary system composed of a 2.50 $M_{\sun}$ primary star coupled with a 2.38 $M_{\sun}$ secondary. To explore different evolutionary scenarios, we performed the estimation at three different times: when the primary is at the end of the central helium burning, when it is at the bottom of the RGB, and when it is in the helium core burning phase. 
The Monte Carlo simulations have been carried out for two typical values of accuracy on the mass determination,  that is, 1\% and 0.1\%.
}
 {Adopting typical observational uncertainties, we found that the recovered age and overshooting efficiency are biased towards low values in all three scenarios. For an uncertainty on the masses of 1\%, the underestimation is particularly relevant for a primary in the central helium burning stage, reaching $-8.5\%$ in age and $-0.04$ ($-25\%$ relative error) in the overshooting parameter $\beta$. In the other scenarios, an undervaluation of the age by about 4\% occurs. A large variability in the fitted values between Monte Carlo simulations was found: for an individual system calibration, the value of the overshooting parameter can vary from $\beta = 0.0$ to $\beta = 0.26$.  
 When adopting a 0.1\% error on the masses, the biases remain nearly unchanged but the global variability is suppressed by a factor of about two.      
 We also explored the effect of a systematic discrepancy between the artificial systems and the model grid by accounting for an offset in the effective temperature of the stars by $\pm 150$ K. For a mass error of 1\% the overshooting parameter is largely biased towards the edges of the explored range, while for the lower mass uncertainty it is basically unconstrained from 0.0 to 0.2. 
 We also evaluate the possibility of individually recovering the $\beta$ value for both binary stars. We found that this is impossible for a primary near to central hydrogen exhaustion owing to huge biases for the primary star of $+0.14$ (90\% relative error), while in the other cases the fitted $\beta$ are consistent, but always biased by about $-0.04$ ($-25\%$ relative error). 
 Finally, the possibility to distinguish between models computed with mild overshooting from models with no overshooting was evaluated, resulting in a reassuring power of distinction greater than 80\%. However, the scenario with a primary in the central helium burning was a notable exception, showing a power of distinction lower than 5\%.       
 }
{}


   \maketitle

\section{Introduction}\label{sec:intro}

The theoretical effort made in the past decades led to a continuous refinement of the predictions of stellar evolution theory.
However, several mechanisms -- chiefly the lack of a self-consistent treatment of convection in stellar evolutionary codes -- remain poorly understood. As a consequence, 
the dimension of the convective core cannot be accurately predicted, and its extent besides the classical Schwarzschild border is obtained in terms of the pressure scale height $H_{\rm p}$: $l_{\rm ov} = \beta H_{\rm p}$, where $\beta$ is a free parameter which needs to be calibrated against observational data. 

To this regard, the importance of double-lined eclipsing binary systems --  stars for which a direct measurement of the mass is achievable -- is well recognised in the literature \citep[see, among many,][]{Andersen1991,Torres2010}. Improvement in observational techniques has made it possible to determine masses, as well as radii, of these objects with a precision of a few percentage points or better \citep[see e.g.][]{Clausen2008,Pavlovski2014}. 
It is therefore natural to look at binary systems as a way of obtaining an empirical calibration of the core overshooting parameter with respect to the stellar mass.
Indeed, an accurate prediction of the extension of the convective core is of paramount importance, because its dimension has a profound impact on stellar evolution and thus age determination. 

Several attempts at an empirical calibration of the overshooting parameter can be found in the literature \citep[e.g.][]{Andersen1991,Ribas2000,Claret2007,Meng2014,Stancliffe2015,Deheuvels2016, Claret2016, Claret2017} from different binary systems. 
Following the procedure first adopted in \citet{Andersen1991}, these studies rely on fitting a set of stellar tracks (or isochrones) to the observational data in order to recover the best overshooting parameter compatible with a coeval solution.
The results of these studies are quite contradictory:  \citet{Meng2014} and \citet{Stancliffe2015} suggest no trend in the fitted overshooting efficiency with the stellar mass, in contrast with the finding by \citet{Ribas2000}, \citet{Claret2007}, and \citet{Claret2016}. However, a meaningful comparison among these studies is quite problematic because they adopt different approaches to estimate the overshooting efficiency. First of all, not all of them include in the analysis the uncertainty affecting stellar mass estimates, which can have a seizable effect on the calibration \citep[see the analysis in ][]{TZFor}. A second difference is that not all the employed stellar models include the central helium burning stage, which can drastically alter the estimates for some of the considered systems. Thirdly, the treatment of the uncertainty associated with the initial chemical composition of the stars, mainly the initial helium abundance which is poorly constrained, is not uniform among the studies. 
Nevertheless, the  
general suggestion that can be deduced is that a low core overshooting parameter $\beta \lesssim 0.2$ seems to be enough to match the observational data. 

In spite of the fact that eclipsing binary stars were commonly used to calibrate stellar parameters, the theoretical foundations of such a procedure have not yet been fully investigated. 
Since the proposal of the method by \citet{Andersen1991}, very few efforts can be found in the
literature to establish the theoretical accuracy of such a calibration, exploring the possible sources of internal bias. 
Indeed, the recent results presented by \citet{overshooting} and \citet{testW} raised some questions about the actual reliability of the procedure, showing a large random variability and a non-negligible bias in the estimated overshooting parameter from  main sequence (MS) stars with masses lower than 1.6 $M_{\sun}$. 
The same difficulties could exist for systems in different mass ranges and evolutionary stages. Therefore, a direct evaluation of the calibration procedure reliability is thus needed. The best way to do this is to build an artificial binary system with the same evolutionary characteristics as the observed one.
The key point to address in this theoretical analysis is whether or not it is possible to accurately recover the parameters adopted to generate the synthetic -- and therefore perfectly known -- binary system, once for the unavoidable errors in the observational data have been  accounted for. This information can be obtained by evaluating the maximum theoretical performance of the calibration procedure, working in an ideal scenario in which the stellar tracks adopted for the fitting are the same as those from which the artificial systems to be fitted are derived. A Monte Carlo procedure can be used to simulate random observational errors in the artificial data and test the reliability of the recovery procedure.

Until recently, such a theoretical analysis of this type was out of reach because it requires a huge computational effort coupled with a statistically robust treatment of the results. 
First of all, it relies on the computation of a large set of stellar tracks, varying the input in a way to cover the possible range of variation of the parameter to be calibrated. Therefore, stellar tracks should be computed on a fine grid of initial metallicity and helium abundances, coupled with different values of the core overshooting parameter in a reasonable uncertainty range. 
This database of tracks can then be adopted for a Monte Carlo recovery of the parameters of a reference system. The results should be subjected to a statistical analysis to establish the accuracy and the precision of the estimated parameters.
   
The aim of this paper is to extend the work presented by \citet{overshooting} for low-mass main sequence stars to higher masses and later evolutionary phases
performing the analysis for a hypothetical binary system composed of two stars of 2.50 $M_{\sun}$ and 2.38 $M_{\sun}$, in the post-MS evolution.

\section{Methods}\label{sec:method}

We generalise the fitting procedure adopted in \citet{overshooting} to
simultaneously account for the observational errors and bias sources in the post-MS evolution. 
The selected test case is a binary system composed of a 2.50 $M_{\sun}$ star coupled with a 2.38 $M_{\sun}$ star. 
 A mass ratio of 1.05 allows us to consider two stars of similar mass -- for comparison, about 40\% of binary systems in the detached eclipsing binary catalogue DEBCat \citep{Southworth2015} have a mass ratio equal to or lower than this -- but being sufficiently different to be located in distinct evolutionary stages. 
Moreover, we can safely assume that both components share the same core overshooting parameter, as they have similar masses.
The target system is characterised by [Fe/H] = 0.0, $Y$ = 0.275, and a fixed convective core overshooting efficiency; a further description is given in Sect.~\ref{sec:grids}.     
The masses and the metallicity of the adopted synthetic system are very close to those of one of the brightest binary stars in the sky, Capella ($\alpha$ Aurigae), a well studied system \citep[see][and references therein]{Torres2015} thus allowing insight into the global accuracy of a fit from this system.

To estimate the reliability of the convective core overshooting efficiency calibration from these binary stars, we developed a two-stage pipeline. The first stage is required for sampling artificial systems from an ideal one, accounting for observational errors, while the second one entails an estimation of stellar parameters for the systems generated in the previous step.

\begin{figure*}
        \centering
        \includegraphics[height=14cm,angle=-90]{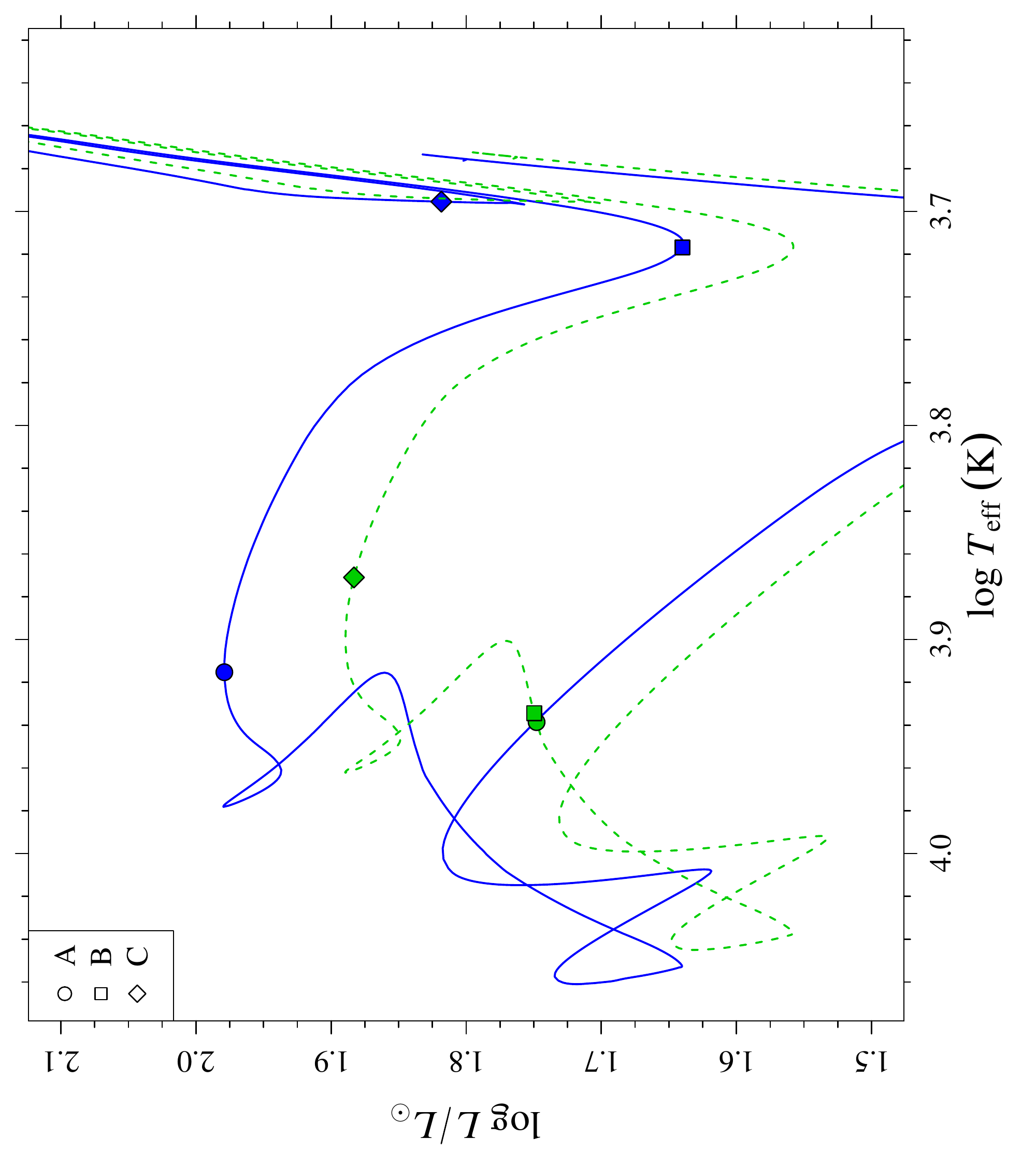}
        \caption{HR diagram for the reference primary $M_1 = 2.50$ $M_{\sun}$ (blue solid line) and secondary $M_2 = 2.38$ $M_{\sun}$ (green dashed line) stars. The three studied scenarios are identified by circles ($A$), squares ($B$), and  diamonds ($C$). The parameters adopted for the stellar evolution are: $M_1 = 2.50$ $M_{\sun}$, $M_2 = 2.38$ $M_{\sun}$, $Z = 0.0134$, $Y = 0.275$, $\beta = 0.16$. 
        }
        \label{fig:trk}
\end{figure*}

In principle, the results could depend on the evolutionary stages of the two stars, therefore we performed the exercise for three configurations to cover different evolutionary stages. 
The positions of the selected systems on the HR diagram are shown in Fig.~\ref{fig:trk}, labelled as $A$, $B$, and $C$.  
Scenario $A$ contains a primary just after the central hydrogen depletion; the secondary is firmly in the MS phase. Scenario $B$ presents the primary at the base of the red giant branch (RGB), while the secondary is nearly in the same position as the previous system. Finally, scenario $C$ has a primary star in the central helium burning stage, when 30\% of the central helium has been burned, and a secondary in the sub giant branch (SGB). This last scenario closely mimics the proposed fit of Capella \citep{Torres2015, Claret2017}, and therefore has particular relevance to judging the robustness of a fit from real stars in these evolutionary stages.

\subsection{Sampling scheme}\label{sec:fittingschema}

We take
${\cal S}_1$ and ${\cal S}_2$ be two stars in a detached binary system, for which the 
observational constraints available are the effective temperatures, the metallicities, the masses, and the radii of the two stars, that is, $\tilde{q}^{{\cal S}_{1,2}} \equiv \{T_{\rm eff, {\cal S}_{1,2}}, {\rm
        [Fe/H]}_{{\cal S}_{1,2}}, 
M_{{\cal S}_{1,2}}, R_{{\cal S}_{1,2}}\}$. We also take $\sigma^{1,2} = \{\sigma(T_{\rm
        eff, {\cal S}_{1,2}}), \sigma({\rm [Fe/H]}_{{\cal S}_{1,2}}), \sigma(M_{{\cal S}_{1,2}}),
\sigma(R_{{\cal S}_{1,2}})\}$ to be the observational uncertainties.
Due to measurement errors, the true values of $\tilde{q}$ remain unknown and one has a proxy for them, the observed values $q$. In the following, we adopt uncertainties of 100 K in the $T_{\rm eff}$, 0.1 dex in [Fe/H], and 0.5\% in the radii. We explored two different uncertainty scenarios for the mass values: 1\% and 0.1\%.
 
As usual, we assume that the density of probability of an observation $q$ follows a multivariate Gaussian distribution with mean $\mu = \{\tilde{q}^{S_1}, \tilde{q}^{S_2}\}$ and covariance matrix $\Sigma$.
Since the observationally inferred values
of a given physical quantity for the two binary components
are correlated, it would be unsafe to adopt a diagonal covariance matrix. Off-diagonal elements computed adopting sensible correlation coefficients should be included in
$\Sigma$ whenever a realistic noise has to be simulated.
We assume a correlation of 0.95 between
the primary and secondary effective temperatures, and 0.95
between the metallicities of the two stars. Regarding masses and radii correlation, we
set the former at 0.8 and the latter at $-0.9$, typical values for this class of stars \citep[see][and references therein]{binary}. 

The first step of our analysis involves the generation of $N_1 = 500$ artificial binary systems from each of the selected evolutionary stages $A$, $B$, and $C$. These artificial systems were generated by sampling from the multivariate Gaussian distribution specified above to account for several possible "observations" of the same binary. The number of systems $N_1$ was established by a trial-and-error procedure, such that it is large enough to ensure the stability of the derived results.

\subsection{Fitting procedure}\label{sec:fittingML}

The procedure followed to estimate the stellar parameters is based on a modified SCEPtER pipeline\footnote{Publicly available on CRAN: \url{http://CRAN.R-project.org/package=SCEPtER}, \url{http://CRAN.R-project.org/package=SCEPtERbinary}}, a well tested grid-based maximum likelihood technique, which has been adopted in the past for single stars \citep{scepter1,eta,bulge}, and for binary systems \citep{binary}. Briefly, the procedure computes a likelihood value for all the grid points and provides estimates of the parameters of interest (age, initial helium abundance, initial metallicity, core overshooting parameter) by averaging the values of all the models with a likelihood greater than the 95\% of the maximum value.

The estimation is performed for each of the $N_1$ artificially generated  systems.
 For each point $j$ on the estimation grid of stellar models, 
we define $q^{j} \equiv \{T_{{\rm eff}, j}, {\rm [Fe/H]}_{j}, M_{j},
R_{j}\}$. 
Letting $ {{\cal L}^{1,2}}_j$ be the single-star likelihood functions defined as
\begin{equation}
        {{\cal L}^{1,2}}_j = \left( \prod_{i=1}^4 \frac{1}{\sqrt{2 \pi}
                \sigma^{1,2}_i} \right) 
        \times \exp \left( -\frac{\chi_{1,2}^2}{2} \right)
        \label{eq:lik}
        ,\end{equation}
where
\begin{equation}
        \chi^2_{1,2} = \sum_{i=1}^4 \left( \frac{q_i^{{\cal S}_{1,2}} -
                q_i^j}{\sigma_i} \right)^2, 
        \label{eq:chi2}
\end{equation}
the joint likelihood  ${\cal \tilde L}$ of the system is computed as the product of the single star likelihood functions.
In the computation of the likelihood functions we consider only the grid points  
within $3 \sigma$ of all the variables from the observational constraints.
The pipeline provides the parameter estimates both for the individual components and for 
the  whole system. In the former case, the fits for the two stars are obtained independently,
while in the latter case, the algorithm imposes that the two members must have a common age (with a tolerance of 1 Myr), identical initial helium abundance and initial metallicity, and a  common overshooting efficiency parameter. In Sect.~\ref{sec:ov} we analyse the results obtained relaxing the latter constraint to evaluate the possibility to recover the core overshooting parameter for both system components independently.

The estimation process relies on stellar models spanning a wide range of evolutionary phases with very different time scales; therefore the time step between consecutive points is far from being uniform. As a consequence, grid based estimates can be biased towards more densely represented  phases. To obtain a sensible estimate of the density function of the age of the system and of the best overshooting parameter consistent with the data, we adopt a weighted approach \citep[see e.g.][]{Jorgensen2005, eta}. Each point in the grid is weighted with the evolutionary time step around it and this weight is inserted as multiplicative factor in Eq.~(\ref{eq:lik}). This procedure intrinsically favours lower helium abundance and higher metallicity models because their evolution is slower. Therefore we allow for a normalization at track level, so that the highest weight for each track is equal to one. However, the impact on the result of such normalization is almost negligible.      

In addition to the estimated values for the stellar parameters it is fundamental to evaluate the associated errors.
These errors are obtained by means of Monte Carlo simulations. 
We generate again $N_2 = 500$ artificial binary systems, sampling from a multivariate Gaussian distribution with mean $\mu = \{q^{S_1}, q^{S_2}\}$ (therefore centred on the individual $N_1$ artificial systems) and covariance matrix $\Sigma$. 
For each of these $N_2$ systems, we repeat the parameter estimates. The errors on the age and convective core overshooting efficiency are assessed by computing the 16th and the 84th quantiles of their marginalized distributions.

At the end of these simulation and estimation steps, we obtained $N_1 \times N_2 = 250\,000$ estimates of stellar parameters for each of the configurations, $A$, $B$, and $C,$ to be subjected to statistical analysis.

\subsection{Stellar model grid}
\label{sec:grids}

The estimation procedure described above requires a grid of stellar models, sufficiently extended so as to cover the whole parameter space. To this purpose, we computed a grid of stellar models by means of the FRANEC code \citep{scilla2008, Tognelli2011}, in the same
configuration as was adopted to compute the Pisa Stellar
Evolution Data Base\footnote{\url{http://astro.df.unipi.it/stellar-models/}} 
for low-mass stars \citep{database2012}. We adopted the solar heavy-element mixture by \citet{AGSS09}, the solar-scaled mixing-length parameter $\alpha_{\rm
ml} = 1.74$, and we neglect microscopic diffusion; due to the fast evolution in considered mass range, microscopic diffusion affects the evolution in a negligible way.
The reference scenario does not account for radiative acceleration, rotational mixing, and magnetic field, which are affected by somewhat large uncertainties. Furthermore, all the models were computed neglecting other input physics and parameter uncertainties, such as the effects of mass-loss, the uncertainties on the opacities, and the equation of state, which are expected to have an impact on the evolution time scale and track path \citep[see e.g.][]{incertezze1, Stancliffe2015}. Although these uncertainties would alter the estimates from real-world objects, their neglect is  expected to only mildly influence the results presented here. In fact, the present study deals with differential effects in an ideal configuration where stellar models and artificial observations perfectly agree with respect to the adopted input physics.

The overshooting region, that is, the 
extension of the mixed region beyond the Schwarzschild border, was parametrized  in terms of the pressure scale height $H_{\rm p}$: $l_{\rm ov} = \beta H_{\rm p}$. An instantaneous mixing is adopted in the overshooting region. Furthermore, the radiative temperature gradient is adopted in this region.
The actual meaning of the $\beta$ parameter is strictly linked to the previous choices, and other overshooting schemes exist in the literature \citep[see e.g.][]{Herwig2000,Zhang2013,Viallet2015,Gabriel2017}. Thus a direct comparison of values presented in different researches is not fully meaningful, because the same $\beta$ value can impact the stellar model
evolution in different ways. A more informative comparison would be that of the extent of the convective core (if present). However,  the simulations presented here adopt the same overshooting mechanism, and therefore the comparison of the $\beta$ values is theoretically justified.

Semi-convection during the central He-burning phase 
\citep{castellani1971} was treated according to the algorithm
described in \citet{castellani1985}, resulting in no additional free parameters to be tuned. Breathing pulses were suppressed \citep{castellani1985, cassisi2001} as suggested by \citet{caputo1989}.  
Further details on the stellar models are fully
described in \citet{cefeidi,eta,binary} and references therein.

We computed a grid of models in the range [2.26; 2.62] $M_{\sun}$, with a step of 0.01 $M_{\sun}$, and the 
initial metallicity interval $-0.4 {\rm dex} \leq$ [Fe/H] $\leq 0.4$ dex, with
a step of 0.05 dex. 
For each metallicity we computed models for nine different values of the initial helium abundance by following the 
linear relation $Y = Y_p+\frac{\Delta Y}{\Delta Z} Z$,
with the primordial abundance $Y_p = 0.2485$ from WMAP
\citep{peimbert07a,peimbert07b} and with a helium-to-metal enrichment ratio $\Delta Y/\Delta Z$
from 1 to 3 with a step of 0.25 \citep{gennaro10}, $\Delta Y/\Delta Z = 2.0$ corresponding to the reference value for the synthetic system. Ultimately, the grid span a set of 153 different initial chemical compositions. For each mass, metallicity and initial helium abundance, we computed models for 18 values of the core overshooting parameter $\beta$ in the range
[0.00; 0.30] with a step of 0.02. The reference value of $\beta$  was chosen to be 0.16.

\section{Results}
\label{sec:results}

The recovery of stellar parameters was performed twice for each of the three scenarios, $A$, $B$, and $C$, 
by taking into account the two adopted uncertainties on the mass of the stars. The 1\% error in mass represents a typical observational goal, nowadays often achieved. A second scenario (labelled $A_M$, $B_M$, $C_M$) adopts an uncertainty of 0.1\% in the masses, achievable in the best cases \citep[e.g.][]{Gallenne2015,Kirkby-Kent2016}.   

The results of the six Monte Carlo simulations are shown in Table.~\ref{tab:mainres} and in Figs.~\ref{fig:OV-age} and \ref{fig:OV-Y}.
The analysis allows us to address some specific questions about the global and individual reliability of the calibration. 

As explained in Sect.~\ref{sec:fittingschema}, the global results were obtained by merging together the  $N_1 \times N_2 = 250\,000$ results for each configuration and offer an overall picture of the expected parameters from the estimation procedure. 
In Sect.~\ref{sec:results-bias} we discuss the biases affecting the estimated system ages and the calibrated overshooting parameter.

Besides the median value of each parameter, which provides an insight into the global expected biases, the most important information is the range spanned by the recovered values due to the observational errors. A large variability in the recovered parameters implies that different artificial perturbations of the same system -- that is, a different random offset from the unobservable true values --  can lead to largely different calibrations. The variability on the results owing to the random observational errors
and its impact on the calibration of a single system is presented in Sect.~\ref{sec:results-var}.

Finally in Sect.~\ref{sec:DT} some results obtained assuming a systematic mismatch between synthetic data and recovery grid in the effective temperature are presented. This result can help to better judge the reliability of the recovery procedure for real-world objects, for which a perfect agreement with stellar model computations is unrealistic.    

\subsection{Bias on the recovered age and parameters}
\label{sec:results-bias}

\begin{figure*}
        \centering
        \includegraphics[height=18cm,angle=-90]{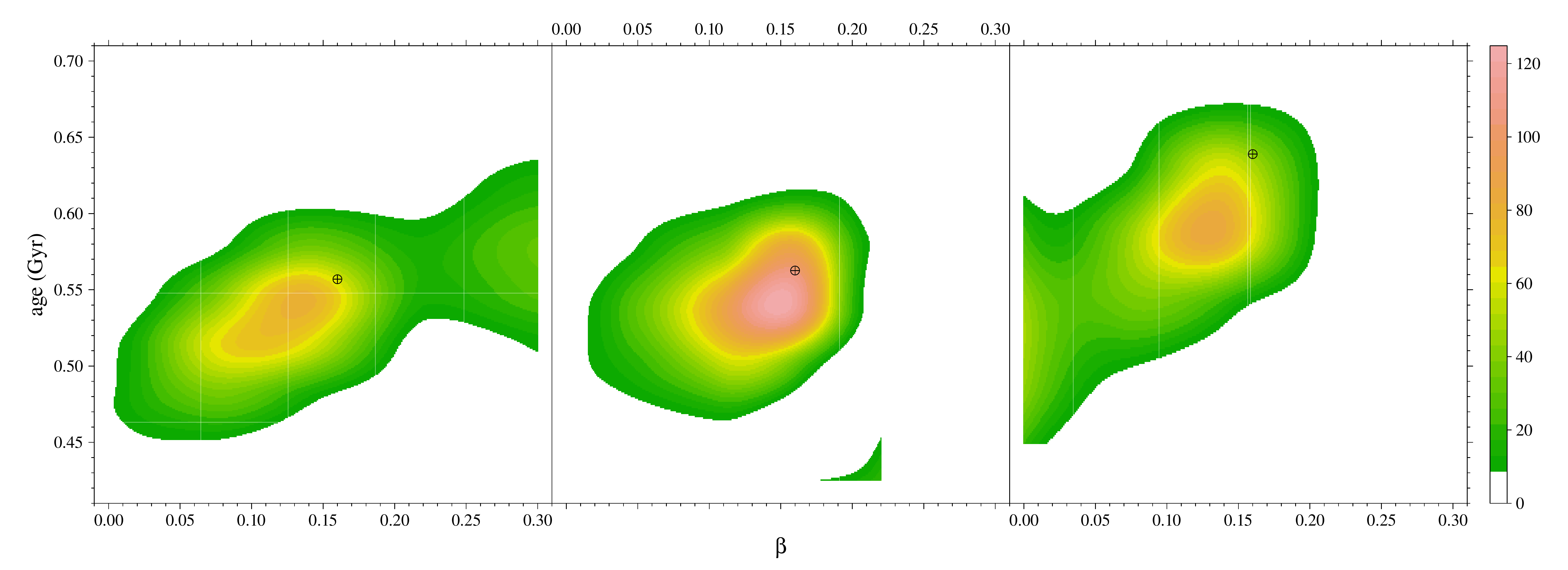}

        \includegraphics[height=18cm,angle=-90]{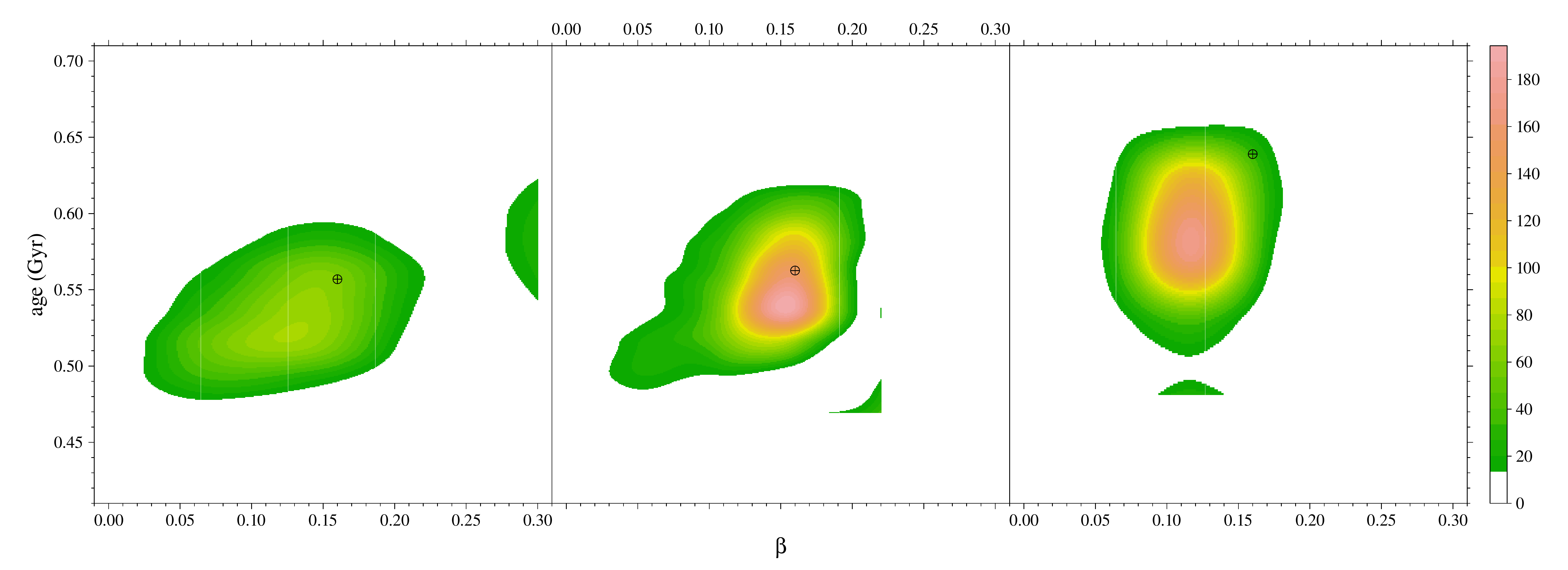}
        \caption{{\it Top row}: {\it left} Bi-dimensional probability density in the $\beta$ vs. age plane, marginalized with respect to initial helium and metallicity for scenario $A$. {\it Middle}: As in the {\it left} panel but for scenario $B$. {\it Right}: As in the {\it left} panel but for scenario $C$. {\it Bottom row}: As in the {\it top row} but for scenarios $A_M$, $B_M$, and $C_M$. The crosses mark  the reference $\beta$ and age values for all six cases. The little island of non-null density in the middle panel of the top row, and those in the middle and right panels of the bottom row, are numerical artefacts.}
        \label{fig:OV-age}
\end{figure*}

\begin{figure*}
        \centering
        \includegraphics[height=18cm,angle=-90]{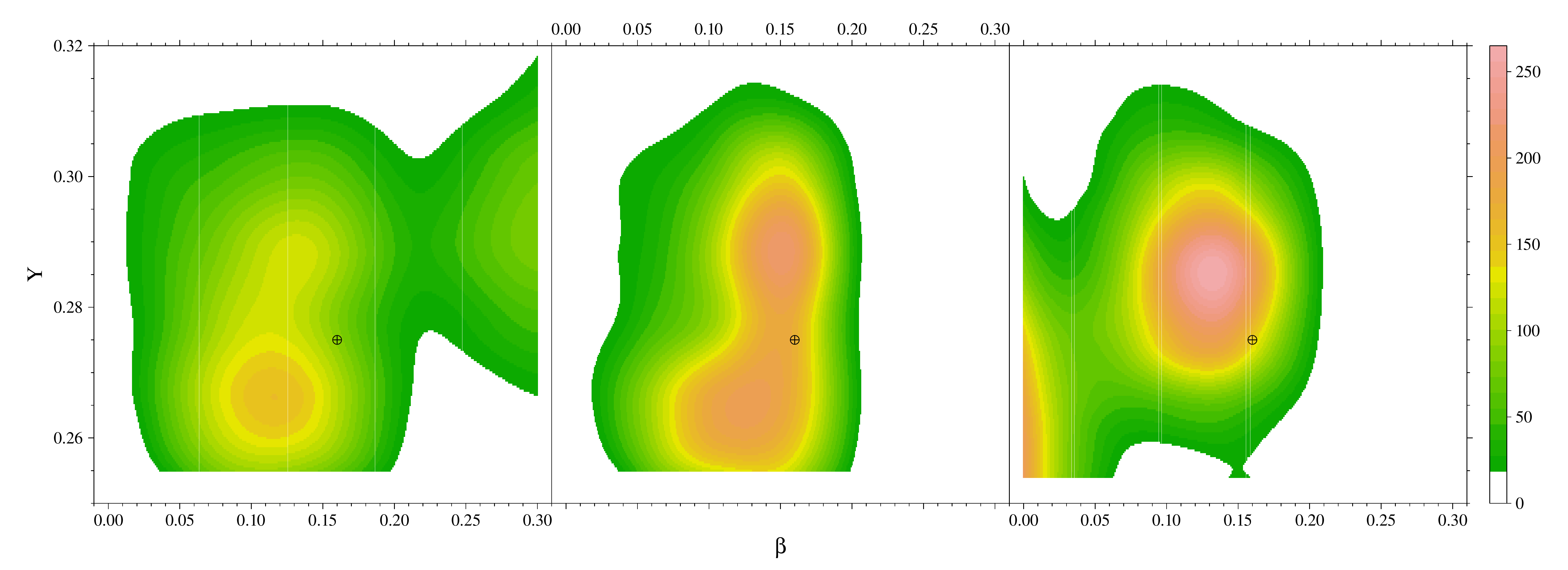}

        \includegraphics[height=18cm,angle=-90]{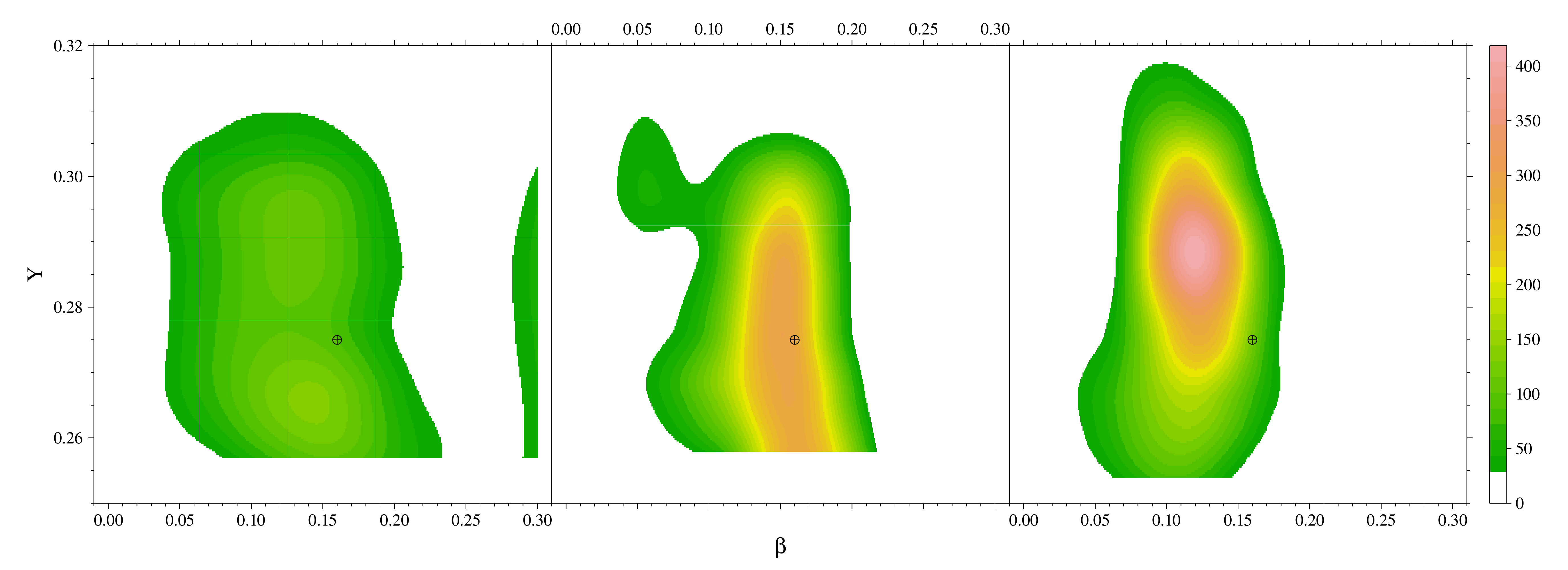}
        \caption{{\it Top row}: {\it left} Bi-dimensional density of probability in the $\beta$ vs. initial helium abundance plane, marginalized with respect to age and metallicity for scenario $A$. {\it Middle}: As in the {\it left} panel but for scenario $B$. {\it Right}: As in the {\it left} panel but for scenario $C$. {\it Bottom row}: As in the {\it top row}, but for scenarios $A_M$, $B_M$, and $C_M$. The crosses mark the position of the reference $\beta$ and age for all six cases.}
        \label{fig:OV-Y}
\end{figure*}

The median value of each estimated parameter is reported in Table~\ref{tab:mainres}, while the errors correspond to the 16th and the 84th quantiles of the marginalized distributions. The same information is presented in Figs.~\ref{fig:OV-age} and \ref{fig:OV-Y} by means of bi-dimensional densities of probability in the age versus $\beta$  and initial helium abundance $Y$ versus $\beta$ planes, marginalized with respect to all the other variables. The crosses indicate the true $\beta$ and age values (Fig.~\ref{fig:OV-age}) and $\beta$ and $Y$ values (Fig.~\ref{fig:OV-Y}) adopted in the simulations.  

\begin{table*}[ht]
        \caption{Estimated stellar parameters from Monte Carlo simulations for the standard ($A$, $B$, $C$) and the more precise scenarios ($A_M$, $B_M$ and $C_M$) described in the text.}
        \label{tab:mainres}
        \centering
        \begin{tabular}{lccccc|cccc}
                \hline\hline
                Scenario & Reference age & Age & Bias & $\beta$ & $Y$ &\multicolumn{2}{c}{Age (Gyr)}&\multicolumn{2}{c}{$\beta$}\\
                & (Gyr) & (Gyr) & (\%) &&& $\sigma$ & $\sigma_g$ & $\sigma$ & $\sigma_g$ \\ 
                \hline
                $A$ & 0.557 & 0.535$^{+0.034}_{-0.032}$ & -3.9 & 0.13$^{+0.11}_{-0.05}$ & 0.281$^{+0.016}_{-0.018}$ & 0.029 & 0.018 &  0.063 & 0.037  \\ 
                $B$ & 0.563 & 0.541$^{+0.027}_{-0.028}$ & -3.8 & 0.13$^{+0.03}_{-0.05}$ & 0.277$^{+0.018}_{-0.015}$ & 0.023 & 0.015 &  0.034 & 0.021  \\ 
                $C$ & 0.639 & 0.585$^{+0.032}_{-0.043}$ & -8.5 & 0.12$^{+0.02}_{-0.08}$ & 0.283$^{+0.012}_{-0.013}$ & 0.034 & 0.021 &  0.040 & 0.027  \\ 
                $A_M$ & 0.557 & 0.538$\pm 0.027$ & -3.4 & 0.14$^{+0.04}_{-0.06}$ & 0.275$^{+0.020}_{-0.012}$ & 0.024 & 0.012 &  0.042 & 0.021  \\ 
                $B_M$ & 0.563 & 0.544$^{+0.042}_{-0.025}$ & -3.3 & 0.14$^{+0.02}_{-0.04}$ & 0.278$^{+0.017}_{-0.015}$ & 0.024 & 0.012 &  0.029 & 0.014  \\ 
                $C_M$ & 0.639 & 0.592$^{+0.016}_{-0.022}$ & -7.3 & 0.12$\pm 0.02$ & 0.285$^{+0.010}_{-0.019}$ & 0.024 & 0.010 &  0.026 & 0.010  \\
                \hline
        \end{tabular}
        \tablefoot{The columns contain: the label of the system to be analysed; the reference age of the system; the estimated age with its $1 \sigma$ error; the relative bias in the estimated age; the estimated overshooting parameter $\beta$ with its $1 \sigma$ error; the estimated initial helium abundance with its $1 \sigma$ error; the two error components of $\sigma$ and $\sigma_g$ for the age;  the two error components of $\sigma$ and $\sigma_g$ for the overshooting (see Sect.~\ref{sec:results-var}). The reference values of overshooting parameter and initial helium abundance are $\beta = 0.16$ and $Y = 0.275$.}
\end{table*}

The estimated age and $\beta$ are biased in all the cases towards lower values with respect to the true ones; while for the scenarios $A$ and $B$ the bias in the age is of about $-4\%$, it is as high as $-8.5\%$ for the more advanced evolutionary stage scenario $C$. These biases in the age are induced by those in the estimated $\beta$ and $Y$. Both these parameters modify the evolutionary time scale and HR position of the stars, but in different ways. As an example, a $\beta$ variation affects the model HR position only in the overall contraction region and afterwards, while a change in the initial helium content is apparent from the beginning of the evolution, as are other ingredients. Regarding the evolutionary time scale, the effect of a variation in the core  overshooting parameter reaches a maximum at about 80\% of the MS lifetime and diminishes afterwards until the central hydrogen exhaustion, while the influence of the initial helium content change is more uniform along the evolution.

In particular, the estimated $\beta$ for scenario $C$ is $0.12\pm 0.02$, that is, $-0.04$ lower than the reference value (i.e. an underestimation of about 25\%), leading to a lower model luminosity with respect to the reference value, which is compensated by an overestimation of the initial helium content by about $0.01$.
These biases are due to the morphology of the grid that is, how the tracks evolve in the hyperspace of parameters adopted for the fitting. As it occurs for different evolutionary stages and masses \citep{bulge, binary}, the tracks at different overshooting and initial helium abundance do not evolve in parallel. It happens that tracks move closer and closer together as the overshooting parameter diminishes, while the opposite is true when $\beta$ increases. Therefore a sample grid point is surrounded by more points at lower $\beta$.    
Hence the likelihood of selecting a track with a lower $\beta$ is higher than that of selecting a higher $\beta$, leading to the observed biases. As a consequence, the effect is not a characteristic of the specific estimation technique adopted in the paper but has a general scope of applicability. This point was also confirmed by means of a direct simulation, adopting in the recovery step -- for scenario $C$ -- only tracks with $\beta = 0.16$ (i.e. the reference value). In this case the bias in age showed a drastic decrease by more than one order of magnitude, from $-8.5\%$ to $-0.7\%$;
once the overshooting parameter is fixed, the initial helium is estimated almost perfectly thus leading to an accurate age prediction. The same behaviour occurs adopting in the recovery only models with the same value of $\Delta Y/\Delta Z$ as the reference one (i.e. $2.0$). In fact the median value of the initial helium provided by the unconstrained fit is higher than the reference one $Y = 0.275$ (see Table~\ref{tab:mainres}) to compensate the decrease in luminosity due to the lower $\beta$. By hampering the fitting algorithm to adjust the solution by providing a biased  initial helium value, that is, restricting the fitting grid to only models with $\Delta Y/\Delta Z = 2.0$, we obtained an unbiased  recovered $\beta = 0.16^{+0.03}_{-0.06}$ -- which however shows the characteristic tendency towards underestimated values --  and an age of 0.631 Gyr, that is, underestimated by 1.2\% with respect to the true value.

Interestingly, the biases in the parameters do not change appreciably  when adopting a lower mass error. This occurs because the median estimated masses in the large error scenarios are not distorted. Thus reducing the uncertainty can modify the propagated  parameter errors but not their median values.
This result is different from the one discussed in \citet{TZFor}, where we showed how a change in the observational precision of mass determination form 1\% to 0.1\% alters in a drastic way the calibration from the TZ For binary system. The difference stems from the fact that in the current work we adopt in the recovery the same set of stellar models used to generate the synthetic binary systems. Therefore, apart from the simulated observational errors, which are of random origin, we have a perfect agreement between targets and recovery grid. This is certainly not the case whenever a real binary system is analysed, because systematic discrepancies between theoretical models and observed data have to be expected. Therefore the present results should be considered to be affected by the very minimum bias and error allowed by the topology of the stellar tracks in the hyperspace of free parameters.

\begin{figure*}
        \centering
        \includegraphics[height=18cm,angle=-90]{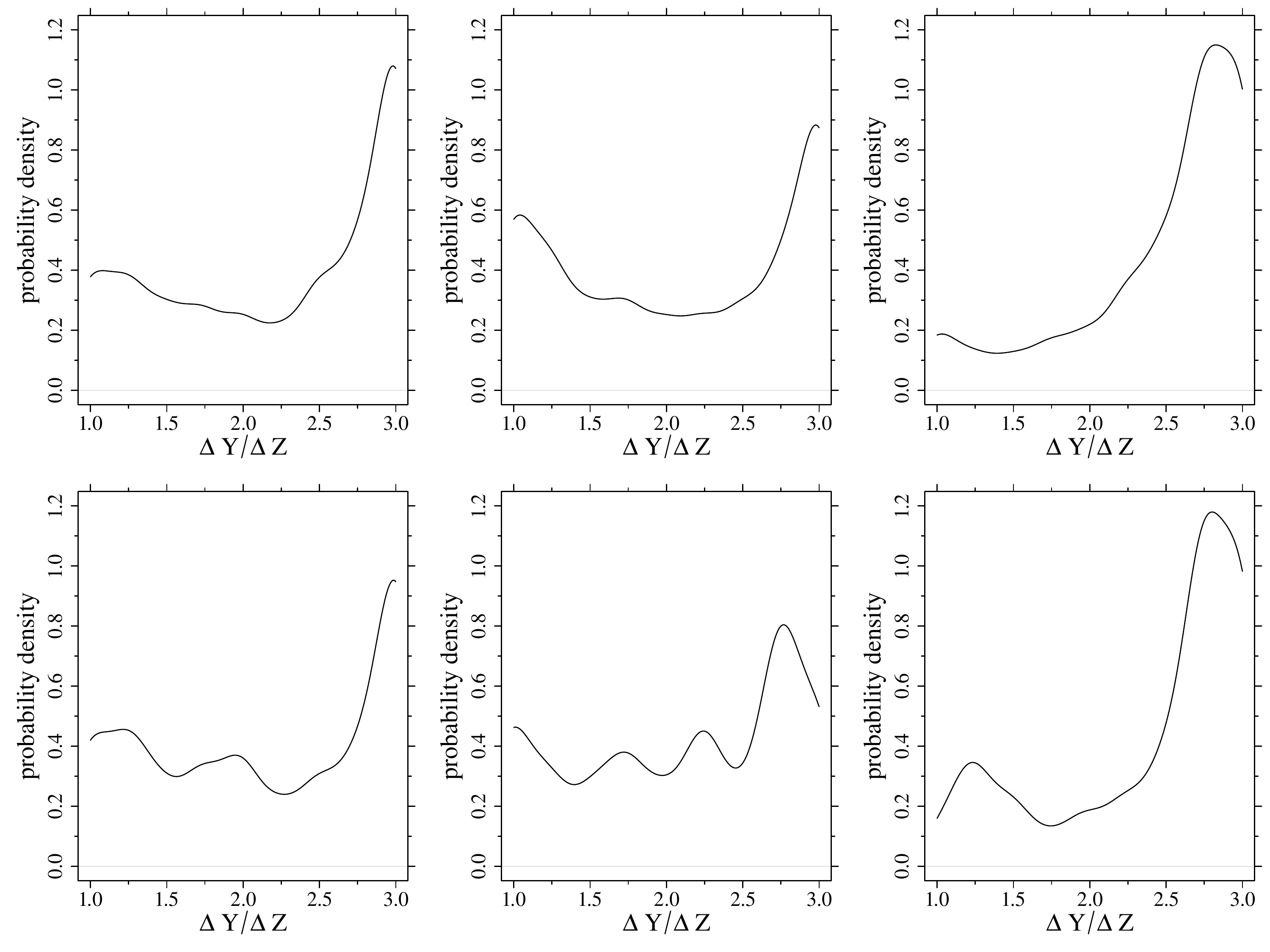}
        \caption{{\it Top row}: {\it left} Estimated density of probability for the helium-to-metal enrichment ratio $\Delta Y/\Delta Z$ for scenario $A$. {\it Middle}: As in the {\it left} panel but for scenario $B$. {\it Right}: As in the {\it left} panel but for scenario $C$. {\it Bottom row}: As in the {\it top row}, but for scenarios $A_M$, $B_M$, and $C_M$.}
        \label{fig:dydz}
\end{figure*}

A relevant cause of the difficulties in the calibration procedure is the uncertainty on the initial helium content. This acts as 
a nuisance parameter that alters the stellar evolution but is not constrained by the observations.
As discussed above, a change in $\beta$ and the initial helium content lead to a bias in the final reconstructed age. Indeed the initial helium abundance -- marginalized with respect to metallicity -- is recovered with large variance and non-negligible bias (see Fig.~\ref{fig:OV-Y}). Interestingly, it also appears that the selected synthetic binary systems do not provide a reliable target for calibration of the helium-to-metal enrichment ratio $\Delta Y/\Delta Z$. 
Figure~\ref{fig:dydz} shows the densities of probability of the recovered $\Delta Y/\Delta Z$ values in the six considered scenarios. While the target value is $\Delta Y/\Delta Z = 2$, the estimates are always distorted with a preference for a higher initial helium abundance. This clearly contributes to the above-mentioned underestimation of age.

\subsection{Random variability of the estimates}
\label{sec:results-var}

Apart from the biases discussed in the previous section, the simulations show a relatively significant dispersion in the age versus $\beta$ plane (Fig.~\ref{fig:OV-age}) and in the initial helium abundance  versus $\beta$ plane (Fig.~\ref{fig:OV-Y}). In particular, the results from scenarios $A$ and $C$ present a large variability in $\beta$: in scenario $A,$  $\beta$ is basically unconstrained in the explored range, in agreement with findings by \citet{overshooting} for MS binary stars. For scenario $A$, the algorithm struggles to predict the initial helium abundance, its probability density being almost flat (Fig.~\ref{fig:OV-Y}).
Scenario $B$ shows the most accurate determination of age and $\beta$, even if the initial helium content is unconstrained  (top middle panel in Fig.~\ref{fig:OV-Y}).
Moreover both  scenarios $A$ and $C$ present multiple individual solutions, despite the very high accuracy adopted for the radii. For scenario $A,$ the most prominent solution locates the primary in the SGB at $\beta \approx 0.13$ (underestimated by approx. 20\%), but also a different solution at $\beta \approx 0.30$ exists, identifying the primary star in the overall contraction phase (see top left panel in  Fig.~\ref{fig:OV-age}). This is not an uncommon result, having already been reported in the literature for some real binary systems \citep{Kirkby-Kent2016,TZFor}.  The same ambiguity exists for scenario $C$; the system can be fitted at $\beta \approx 0.13$ and at $\beta \approx 0.0$, with huge underestimation in the system age by about $15\%$ (see top-right panel in  Fig.~\ref{fig:OV-age}). 
This solution places the primary in the very early helium-burning stages or in the RGB; the secondary is firmly in the SGB phase.   
The presence of multiple solutions is strictly linked to the uncertainty in the stellar masses. For example, in scenario $A$, the estimated masses for the solutions with median $\beta > 0.2$ are $2.50 \pm 0.04$ $M_{\sun}$ and $2.36 \pm 0.03$ $M_{\sun}$ for the primary and secondary stars, showing a relevant offset of $-0.02$ $M_{\sun}$ for the secondary member. For the solution at lower overshooting both the mass estimates are unbiased. 
A possible remedy for these fitting difficulties would come from the availability of asteroseismic observables for such systems \citep{Gaulme2016, Brogaard2018}, which could allow to better identify the evolutionary phase of the stars and lift the degeneracy in the solutions. Indeed, the identification of gravity mode spacings allows a neat distinction of RGB from red clump stars \citep[see e.g.][]{Mosser2011,Bedding2011}.

Similar behaviour can be seen in scenario $C$: restricting to the solutions with median $\beta \leq 0.05,$ the estimated stellar masses are $2.49_{-0.03}^{+0.04}$ $M_{\sun}$ and $2.39_{-0.04}^{+0.03}$ $M_{\sun}$ for the primary and secondary stars, respectively.
As expected, and as shown in \citet{TZFor}, these supplementary solutions disappear in the very precise scenarios $A_M$ and $C_M$, due to the impossibility of exploring the same range of masses of the previous cases.
This is another demonstration of the need for extremely precise observational constraints, particularly on the mass of the stars, when a binary system is adopted for calibration purposes.
It is in fact clear that  a large mass uncertainty  can potentially mask a systematic disagreement between stellar models and observed data, given the freedom of the fitting algorithm in providing solutions for different combinations of masses inside their nominal error.

To highlight how the uncertainty in the observational data can alter the calibration, we present the analyses performed on some of the artificial binary systems resulting from the Gaussian perturbation of their observable constraints. 
This allows us to perceive the huge variability in the maximum likelihood estimates of stellar parameters due to the individual differences between a Monte Carlo simulated system and the theoretical one. Therefore, instead of producing a large bi-dimensional density of probability as before, we focus here on some individual cases under scenarios $A$ and $C$, which showed large variance in $\beta$. This exercise can be helpful to gauge the extent to which a random error in the observational constraints, even assuming no systematic discrepancies between the reconstruction grid and the synthetic systems, can influence the calibration of stellar age and $\beta$.
Each examined case is therefore a single realisation of the $N_1$ perturbations of the observational constraints described in Sect.~\ref{sec:fittingschema}.

\begin{figure*}
        \centering
        \includegraphics[height=9cm,angle=-90]{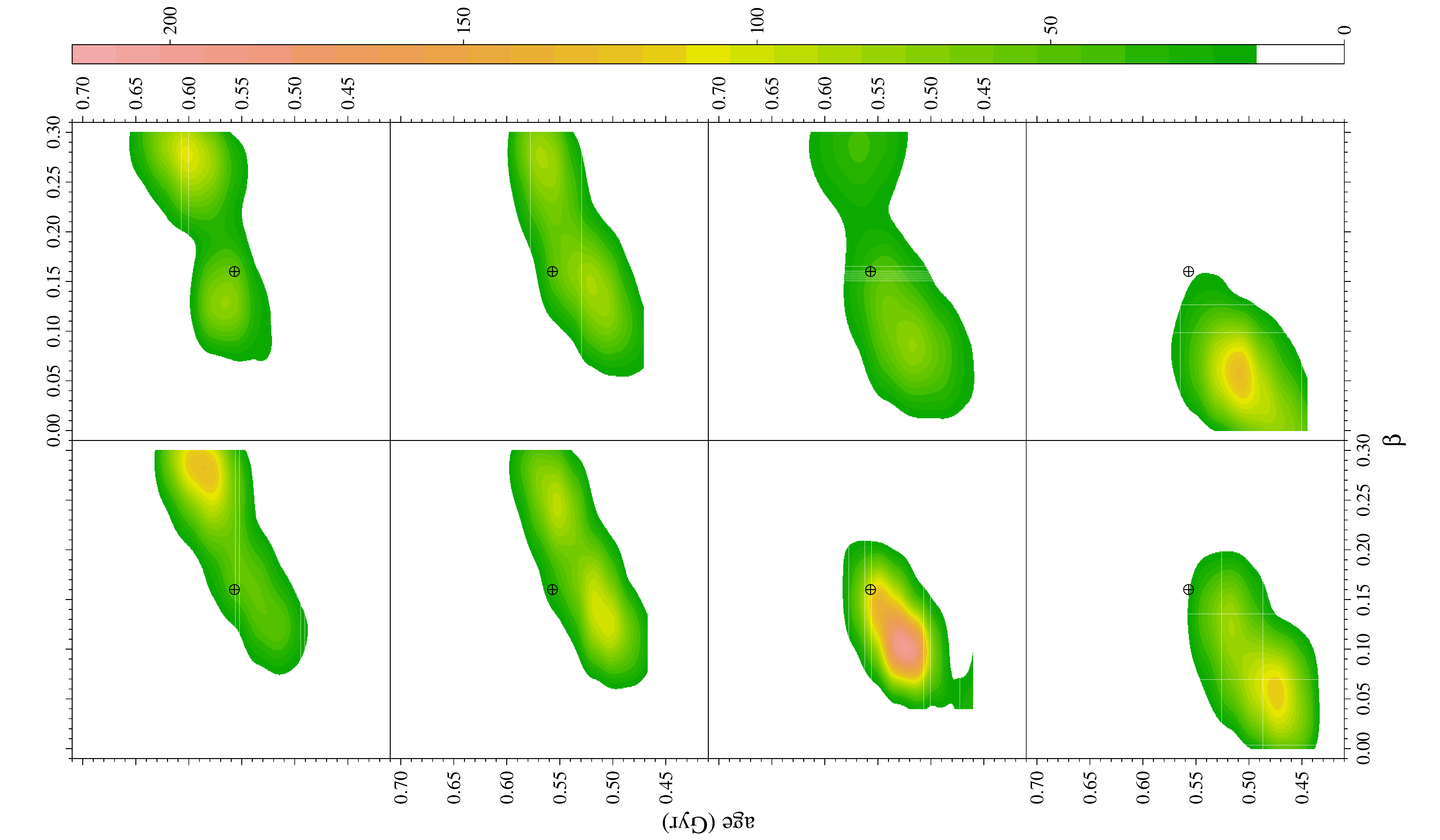}
        \includegraphics[height=9cm,angle=-90]{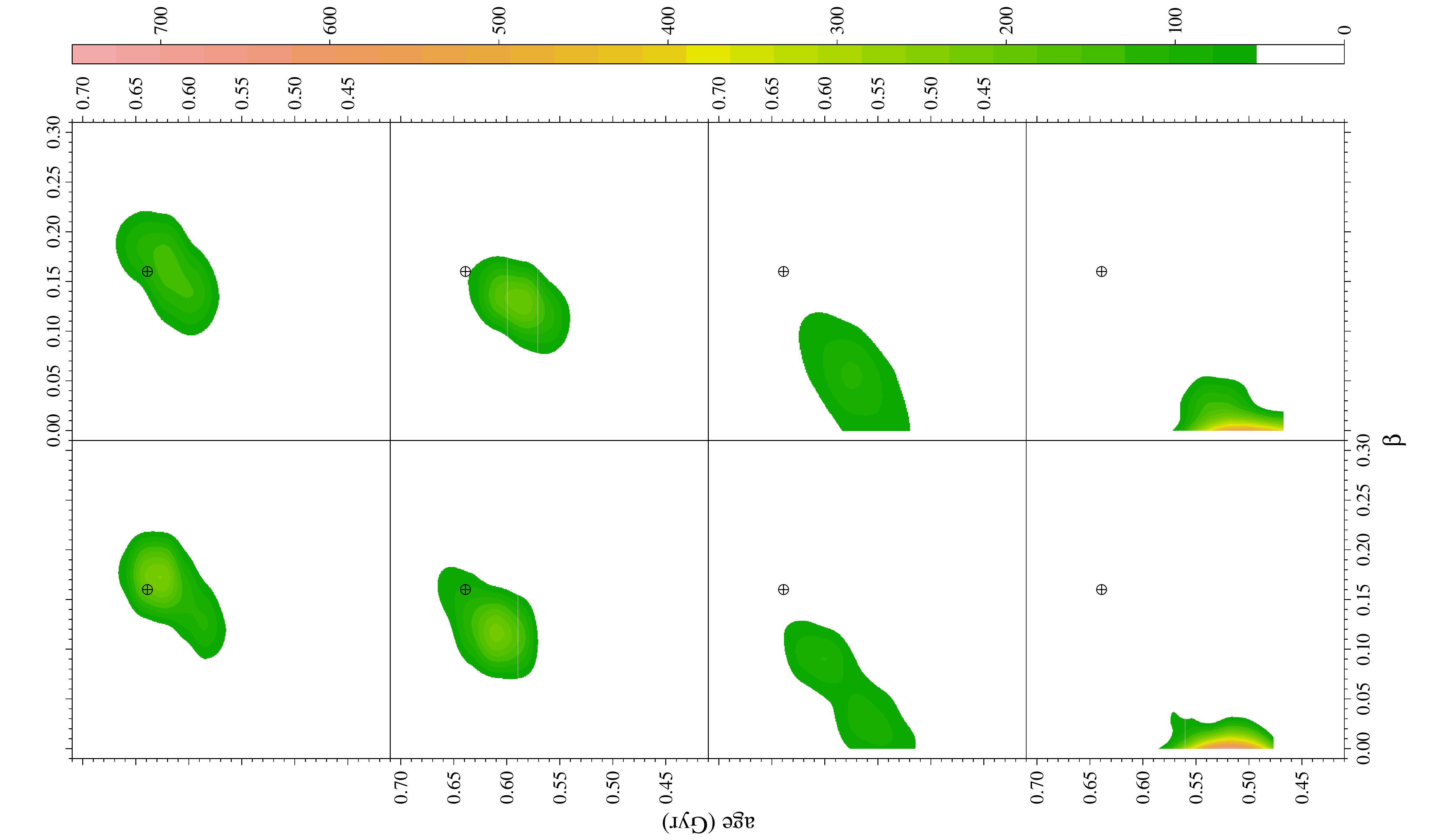}
        \caption{{\it Left} panel, {\it top} row: Two realisations of the bi-dimensional density of probability in the age vs. $\beta$ plane for two artificial systems under scenario $A$, with median estimated $\beta = 0.26$. {\it Second row}: As in the {\it top row} but for median estimated $\beta = 0.18$. {\it Third row}: As in the {\it top row} but for median estimated $\beta = 0.12$. {\it Bottom row}: As in the {\it top row} but for median estimated $\beta = 0.06$. The shown probability densities correspond to eight artificially perturbed systems.
        {\it Right} panel: As in the left panel for scenario $C$; from top to bottom the estimated $\beta$ are 0.16, 0.12, 0.06, 0.0.                          
                The crosses indicate the position of the reference $\beta$ and age values.}
        \label{fig:multiple3-6}
\end{figure*}

Figure~\ref{fig:multiple3-6} shows eight single bi-dimensional densities of probability in the age versus $\beta$ plane in the case of scenario $A$, and eight from scenario $C$. For the first scenario, they were randomly selected to cover the range of fitted $\beta \in$ [0.06; 0.26], while they come from  reconstructed $\beta \in$ [0.0; 0.16] for scenario $C$. 
The left panel of the figure (scenario $A$) shows a clear trend from a solution with a primary preferentially in the SGB phase (top row) to one with a primary in the overall contraction phase (bottom row).
It is clear that different synthetic observational data originating from the same theoretical system can lead to largely different calibrations. 
This situation is even more pronounced in the case of scenario $C$ (right panel in Fig.~\ref{fig:multiple3-6}).

Overall, Figure~\ref{fig:multiple3-6} shows that, in both cases and starting from the same targets, it is possible to obtain both a large underestimation of the $\beta$ value with a concomitant age underestimation, and the inverse: a large $\beta$ with a concomitant age overestimation. 
It is important to note that the cases displayed in the figures do not have an equal probability of occurrence under the Gaussian assumption on the observational errors, as they were selected to cover the whole $\beta$ range.
Their purpose is only to demonstrate the impressive individual variability than can be expected to occur only by chance, even when synthetic targets are sampled from the same grid of models adopted in the recovery.
However, although the extreme perturbations presented in the figure have a low chance of occurrence,  these results urge great caution when relying on an individual calibration, as the latter could be simply a mere by-product of a statistical fluctuation in the observational constraints.  

A better understanding of the importance of the offset from observational and true latent data can be obtained by disentangling the contribution of the different sources of variability in the recovery procedure.  
As a matter of fact, the global variability  in the reconstructed stellar age and parameters arises from two different sources. The first one is the Monte Carlo random component $\sigma$ owing to the reconstruction of the single $N_1$ artificial systems (the so-called within-system variability), itself due to the direct propagation of observational uncertainty on the recovered values for each system. A second source of variability arises from the fact that the artificial $N_1$ systems originate from different perturbations of the unobserved true values; this leads to a difference in the $N_1$ median estimates of age and $\beta$. Thus we have to account for the variability $\sigma_g$ of the median reconstructed parameters from different systems (the so called between-systems variability). 
The first component is straightforwardly evaluable for a real observed system relying on Monte Carlo simulations.  The second one -- which cannot be determined in practice -- originates from the unknown observational bias 
in the observations from the real data. It is clear that the calibration procedure from an object can be considered reliable only when the second source of variability is negligible. In this case, the calibration will result in nearly the same values for whatever unknown shift in the observational data in the range allowed by the observational uncertainties. 

It is possible to gain insight into the relative importance of the two sources of variability
by exploiting a random-effect model \citep{venables2002modern,lme4}, a powerful statistical technique (see Appendix~\ref{app:raneff}).
This information is presented for the age and the $\beta$ parameter in the last four columns of Table~\ref{tab:mainres}.   
It appears that for all the studied cases, the internal variability $\sigma$ is larger than $\sigma_g$ by a factor of 1.5. Hence, the main source of variability is due to the single system reconstructions, but the variability due to the different ways the observations can be biased with respect to true values is only slightly smaller. From the $\sigma_g$ values, it appears that the $\beta$ calibration is affected by a systematic uncertainty of about 0.03 (about 20\% relative error) and 0.015 (about 10\% relative error) in the inaccurate- and accurate-mass scenarios, respectively.    

In summary, the analysis highlights that, even when the same set of stellar models is used in the Monte Carlo sampling and in the reconstruction, the calibration is biased in age and $\beta$. Moreover, different Monte Carlo perturbations of the observations can lead to somewhat different calibrations. Ultimately, it seems that the obtained calibration of the overshooting parameter for a primary star already evolved past the MS should be considered affected by a systematic $1 \sigma$ error of about $\pm 0.03$ ($\pm 20\%$). Instead, the results from scenario $A$ -- where the primary star is just after the central hydrogen exhaustion -- suggest that it is better to avoid a calibration from such a configuration, in agreement with the finding discussed in \citet{overshooting} for MS binary stars. 
For a calibration against a real binary system, the expected situation is even worse, because one cannot be confident that the adopted stellar tracks perfectly describe the reality and that, in some cases, the two stars evolve fully independently; some aspects of this question are explored in Sect.~\ref{sec:DT}.
Therefore, the extreme variations in the first and last rows of Fig.~\ref{fig:multiple3-6}, which arise from large shifts in the observational constraints with respect to the true values, would occur more commonly than in our ideal case.

\subsection{Effect of systematic uncertainties}
\label{sec:DT}

In the previous sections we assumed that the unperturbed artificial systems and the recovery grid perfectly match, and only accounted for Gaussian random perturbations simulating the observational errors. For real data this assumption is obviously overoptimistic and systematic discrepancies must be expected. 

These differences could arise either from incorrect assumptions in the stellar model computations (input physics or chemical abundances) or from observational systematic errors. As an example of the former uncertainty, radiative opacities, EOS, mass-loss efficiency, rotational mixing, and boundary conditions are usually fixed in the computations and the degrees of freedom due to the implemented choices are consequently neglected. Similarly, stellar grids are usually computed  assuming a reference solar composition. 
Regrading observational errors, in particular the absolute calibration of the effective temperature, is still affected by large uncertainties. Although the various works in the literature claim an accuracy of a few tenths of a Kelvin, a direct comparison of these results reveals that differences larger than 100 K are not uncommon \citep[see e.g.][]{Ramirez2005,Schmidt2016}. 
Therefore we performed a Monte Carlo exercise to evaluate the impact of an incorrect calibration of the effective temperature. We artificially perturbed the effective temperature of the synthetic stars, for scenario $C$, by $\pm 150$ K and then reconstructed them assuming an error on the masses of either 1\% or 0.1\%.

\begin{figure*}
        \centering
        \includegraphics[height=18cm,angle=-90]{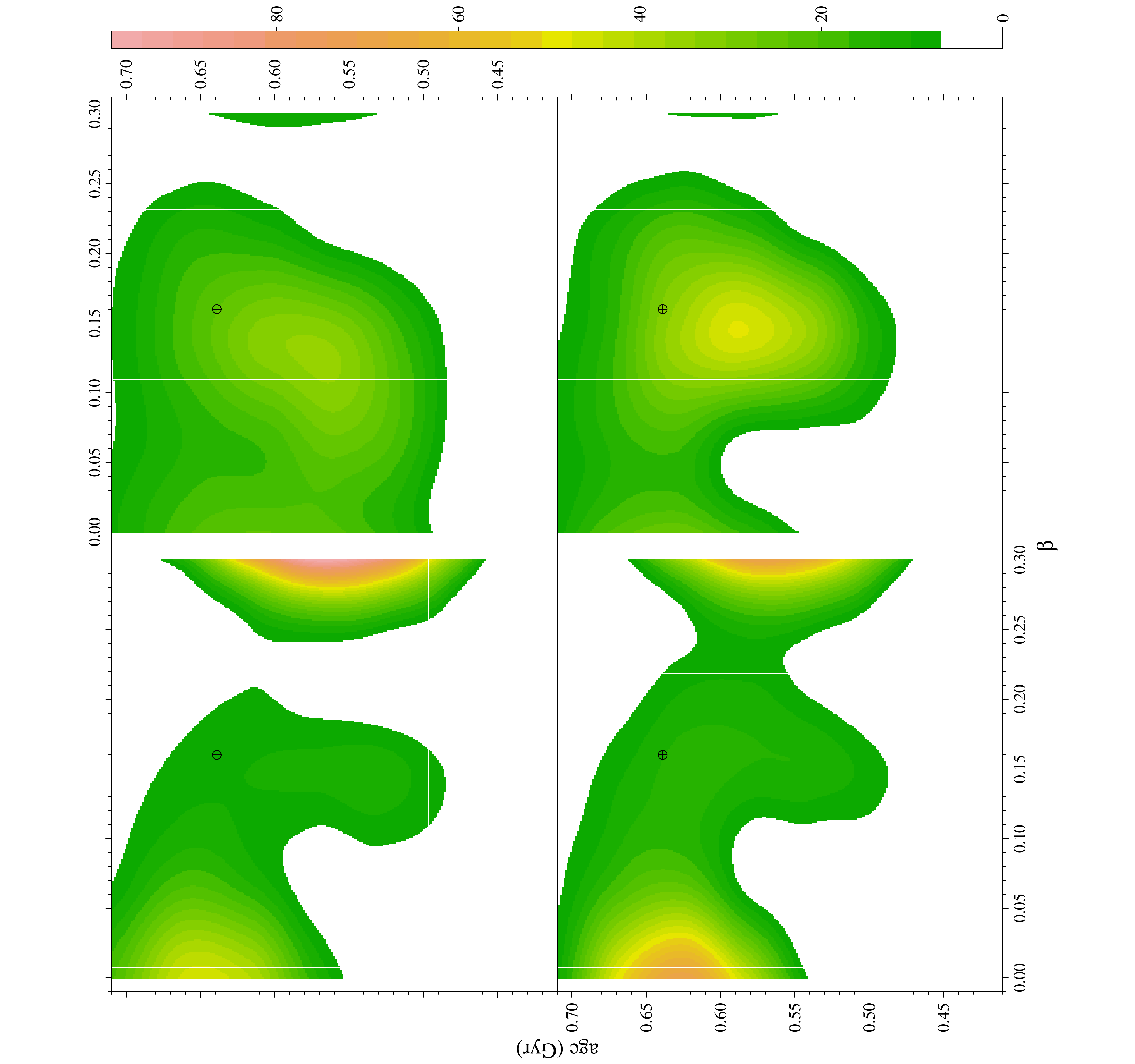}
        \caption{{\it Top} row, {\it left} panel: Bi-dimensional density of probability in the age vs. $\beta$ plane for scenario $C$, adopting a shift in effective temperature of artificial observations of $+150$ K. {\it Right}: Same as in the {\it left} panel, but for a shift of $-150$ K. {\it Bottom} row: Same as in the {\it top} row, but adopting an error on masses of 0.1\%.}
        \label{fig:DT}
\end{figure*}

The results of these simulations are presented in Fig.~\ref{fig:DT}. The left panel in the top row shows the reconstruction of the systems with artificially increased effective temperature. A comparison with the right panel in the top row of Fig.~\ref{fig:OV-age} shows that the main peak of the density is drastically shifted at the grid edge at $\beta \geq 0.25$, with an estimated age of $0.56 \pm 0.03$ Gyr, substantially biased with respect to the target age of 0.64 Gyr.  
Another solution suggests a fit with lower core overshooting efficiency, with a peak in the region $\beta \leq 0.1$ and an estimated age of $0.64 \pm 0.03$ Gyr.
The right panel in the top row  of Fig.~\ref{fig:DT} shows the results for the shift of $-150$ K; in this case the density is almost flat and $\beta$ is unconstrained from 0.0 to 0.2 with  a recovered age of $0.60 \pm 0.06$ Gyr. 

The  bottom row in Fig.~\ref{fig:DT} shows the results assuming an uncertainty on the masses of 0.1\%. For the shift of $+150$ K, the peak at $\beta \geq 0.25$ -- with an age estimate of $0.57^{+0.01}_{-0.02}$ Gyr -- is reduced to the advantage of solutions at $\beta \leq 0.1$ and an age
estimate of $0.63^{+0.03}_{-0.02}$ Gyr.
For the $-150$ K shift, the density is peaked at $\beta = 0.15$, but a large variability is still present; the age estimate is $0.62^{+0.05}_{-0.06}$ Gyr.  
Overall the consequences of different assumptions on the mass errors are mild, suggesting that the displacement of the effective temperature hampers a calibration in the age vs. $\beta$ plane even when very precise determinations of stellar fundamental parameters are available. In fact, the fitting algorithm exploits the indetermination in the initial helium content to find a fit, but converging to biased values.

In summary, by systematically offsetting one observable constraint, the results of the simulations show that the recovered values change with respect to the reference scenario of no systematic offset. This aspect should be considered for the fitting of a real binary system, because many discrepancies between the stellar grid adopted in the recovery and the observed stars are to be expected.  

\subsection{The possibility of discriminating mild from no overshooting models}

In the previous section we focused our attention on one specific reference case with fixed convective core overshooting efficiency $\beta = 0.16$, showing that the recovered overshooting parameter is generally biased towards lower values. 
This behaviour could pose some difficulties when trying to discriminate between models sampled at mild overshooting, as the one discussed above, from models sampled at very low or no overshooting. For the latter, a bias towards overestimated $\beta$ values is expected to occur, because this parameter cannot be negative. Moreover, \citet{overshooting} showed that the biases that affect the reconstruction of a specific scenario are not always representative of those affecting different case studies. It is therefore interesting to directly explore the capability of the recovery procedure to discriminating models sampled at $\beta = 0.0$ from those sampled at $\beta = 0.16$.

\begin{table*}[ht]
\caption{As in Table~\ref{tab:mainres}, but for target binary systems sampled at $\beta = 0.0$.}
\label{tab:mainres-ov0}
\centering
\begin{tabular}{lccccc|cccc}
\hline\hline
Scenario & Reference age & Age & Bias & $\beta$ & $Y$ &\multicolumn{2}{c}{Age (Gyr)}&\multicolumn{2}{c}{$\beta$}\\
& (Gyr) & (Gyr) & (\%) &&& $\sigma$ & $\sigma_g$ & $\sigma$ & $\sigma_g$ \\
\hline
$B$ & 0.501 & 0.506$^{+0.028}_{-0.023}$ & 1.0 &  0.0$1^{+0.02}_{-0.01}$  &
0.272$^{+0.022}_{-0.011}$ & 0.020 & 0.013 & 0.014 & 0.007 \\
$C$ & 0.564 & 0.56$7^{+0.029}_{0.025}$ & 0.6 &  0.02$^{+0.08}_{-0.02}$ &
0.275$^{+ 0.011}_{-0.012}$ & 0.024 & 0.015 & 0.068 & 0.035 \\
$B_M$ & 0.501 & 0.508$^{+0.026}_{0.029}$ & 1.3 &  0.02$\pm 0.02$  &
0.275$^{+ 0.016}_{-0.012}$ & 0.021 & 0.012  & 0.015 & 0.005  \\
$C_M$ & 0.564 & 0.576$^{+0.041}_{0.033}$ & 2.2 &  0.02$^{+0.04}_{-0.02}$  &
0.275$^{+ 0.004}_{-0.011}$ & 0.028 & 0.019  & 0.046 & 0.017  \\
\hline
\end{tabular}
\end{table*}

\begin{figure*}
        \centering
        \includegraphics[height=18cm,angle=-90]{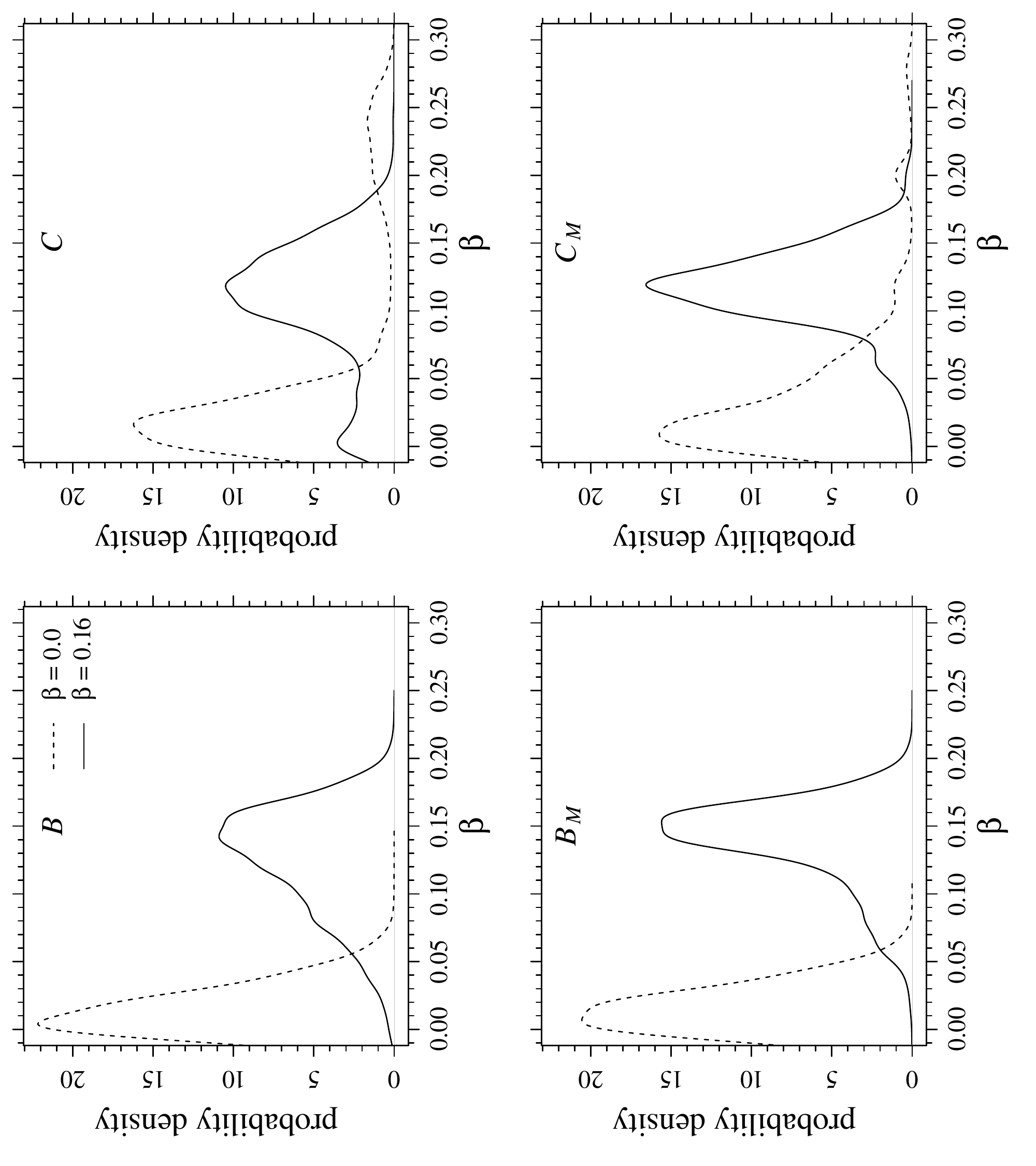}
        \caption{{\it Top} row, {\it left} panel: Marginalized density of probability for the recovered $\beta$ parameter under scenario $B$. The solid line corresponds to models sampled from the grid at $\beta = 0.16$, while the dashed one to models sampled at $\beta = 0.0$. {\it Right} panel: As in the {\it left} panel, but under scenario $C$. {\it Bottom} row: same as in the {\it top} row, but for precise mass scenarios $B_M$ and $C_M$.}
        \label{fig:cfr-ov}
\end{figure*}

To this purpose we repeated the procedure performed above for two test cases mimicking scenarios $B$ (primary at the bottom of the RGB) and $C$ (primary having already burned 30\% of the central helium) discussed in Sect.~\ref{sec:method}, but chosen at $\beta = 0.0$. Scenario $A$ was neglected due to the poor performances of the recovery discussed in the previous sections. The reconstructed age and overshooting efficiency are listed in Table~\ref{tab:mainres-ov0}, while the corresponding marginalized probability densities are shown in Fig.~\ref{fig:cfr-ov}. Some edge effects are apparent. The impossibility to underestimate the value of $\beta$ leads to a modest bias towards higher values. The reconstruction performed under scenarios $C$ and $C_M$ shows a long tail towards higher  $\beta$ values, while -- as occurred for models sampled at $\beta = 0.16$ -- scenario $B$ provides the best results in terms of accuracy. The initial helium value is recovered much better in scenario  $B$  than for models sampled at $\beta = 0.16$, resulting in lower biases in the recovered ages, which are about 1\%. 

From Fig.~\ref{fig:cfr-ov} it is apparent that in most cases the distributions have little overlap, implying that the distinction  between models at $\beta = 0.0$ and $\beta = 0.16$ is quite reliable when no systematic discrepancies between models and synthetic stars exist. Some potential problems arise under scenario $C$ because of the presence of an island of solutions around $\beta = 0.24$ for models sampled at $\beta = 0.0$. This is very similar to the solutions discussed in \citet{TZFor}, and  stems from the possibility to identify the secondary star in the overall contraction and not in the SGB phase, from which it is sampled.

\begin{table}[ht]
        \caption{Powers of the test between models sampled at $\beta = 0.0$ and $\beta = 0.16$.}
        \label{tab:power}
        \centering
        \begin{tabular}{lrrr}
                \hline\hline
                Scenario & 68\% & 90\% & 95\% \\ 
                \hline
                $B$ & 0.981 & 0.948 & 0.948 \\ 
                $C$ & 0.868 & 0.003 & < 0.001 \\ 
                $B_M$ & 0.997 & 0.989 & 0.989 \\ 
                $C_M$ & 0.982 & 0.896 & 0.316 \\ 
                \hline
        \end{tabular}
\tablefoot{The columns contain the power of the test conducted at three different levels, respectively 68\%, 90\% and 95\%.}
\end{table}

To formalize the possibility to differentiate between models sampled at $\beta = 0.0$ and $\beta = 0.16$, it is possible to evaluate, given the marginalized posterior densities, the power of the distinction. Adopting the classical framework of hypotheses comparison \citep{snedecor1989, simar}, we choose as reference hypothesis $H_0$  that a model is sampled at $\beta = 0.0$. 
The alternative hypothesis $H_1$ implies that the model is sampled at $\beta = 0.16$.
Obviously, the two hypotheses can be distinguished well when the distributions of the recovered parameters peak in different $\beta$ ranges and have low variances. The hypothesis comparison  
framework allows a quantitative evaluation of the capability of operating this distinction.
Under the hypothesis $H_0$ we begin by choosing a cut-off $\beta$ value, considered to be "too high" to come from the $\beta$ distribution under $H_0$. Usually this value is set by adopting some reference quantile;  we performed our computation for three different quantiles, namely 68\% ($1 \sigma$), 90\% and 95\% ($2 \sigma$, the value that is more commonly adopted in statistical testing). 
Once these quantiles are evaluated, one should consider how large the portion of the $\beta$ distribution under the hypothesis $H_1$ is for values of $\beta$ larger than the stated quantiles. This is the quantity in which we are interested, and is commonly known as the power of the test. A power above 0.8 is usually considered satisfying. 

Table~\ref{tab:power} lists the power obtained under the different scenarios at the three stated cut-off values. For scenario $B$ the powers are high, well above 0.9.
This result is generally reassuring for the possibility of correctly identifying the need for a non-null convective core overshooting efficiency in the fit. However the very poor power resulting under scenario $C$ (and $C_M$ for the 95\% quantile) suggests that the results of the fit can be, in some cases, drastically misleading. 

This procedure can be further improved by considering the power obtained for the whole set of quantiles $\alpha$ from 0\% to 100\%. The plot of the power of the test versus $1-\alpha$ is a 
well-known statistical diagnostic tool, the  receiver operating characteristic curve (ROC curve) commonly adopted to show the validity of a classifier as the discrimination threshold varies \citep[see e.g.][]{Agresti2013}.
In this framework, the area under the ROC curve, bounded from 0 to 1, is adopted as a global performance of the classifier. A value of 0.5 corresponds to a non-informative classifier, that is, the classification is identical to picking the class at random. Values close to 1 correspond to a very good classifier.   
The evaluation of the areas under the ROC curves resulted in the values of 0.987, 0.767, 0.997, and 0.927, for scenarios $B$, $C$, $B_M$, and $C_M$, respectively. With the exception of scenario $C$, the values are remarkably high.

\section{Allowing independent overshooting efficiencies for the two stars}\label{sec:ov}

The analysis presented in the previous sections adopts, in the reconstruction of the artificial systems, a common overshooting $\beta$ for the two binary stars. Although for the considered systems this hypothesis derives from theoretical considerations,  the stars having identical chemical compositions and very similar masses, a mass ratio of the stars far from one can be justified. Therefore in this section we present an analysis of the internal variability on the individually fitted $\beta$ values that could be expected to occur only owing to random fluctuations. To this aim we relax the constraint of the common overshooting parameter in the recovering algorithm and allow each star to have an individual $\beta$.

\begin{table*}[ht]
\caption{Estimated individual overshooting efficiency for primary and secondary stars in different scenarios and their differences.}
\label{tab:betasingle}  
\centering
        \begin{tabular}{lccccccccc}
                \hline\hline
                Scenario & \multicolumn{3}{c}{$\beta_1$} &  \multicolumn{3}{c}{$\beta_2$} &  \multicolumn{3}{c}{$\beta_1 - \beta_2$} \\
                & $q_{16}$ & $q_{50}$ & $q_{84}$ & $q_{16}$ & $q_{50}$ & $q_{84}$ & $q_{16}$ & $q_{50}$ & $q_{84}$ \\ 
                \hline
                $A$ & 0.20 & 0.30 & 0.30 & 0.12 & 0.14 & 0.18 & 0.04 & 0.14 & 0.17 \\ 
                $B$ & 0.10 & 0.14 & 0.16 & 0.08 & 0.12 & 0.19 & -0.04 & 0.01 & 0.03 \\ 
                $C$ & 0.09 & 0.11 & 0.13 & 0.08 & 0.10 & 0.14 & -0.03 & 0.01 & 0.03 \\ 
                \hline
        \end{tabular}
        \tablefoot{The table shows the 16th, 50th and 84th quantiles of the fitted overshooting efficiency $\beta_1$ (columns 1-3), of the fitted overshooting efficiency $\beta_2$ (columns 4-6), and of their difference (columns 7-9).}
\end{table*}

\begin{figure*}
        \centering
        \includegraphics[height=6cm,angle=-90]{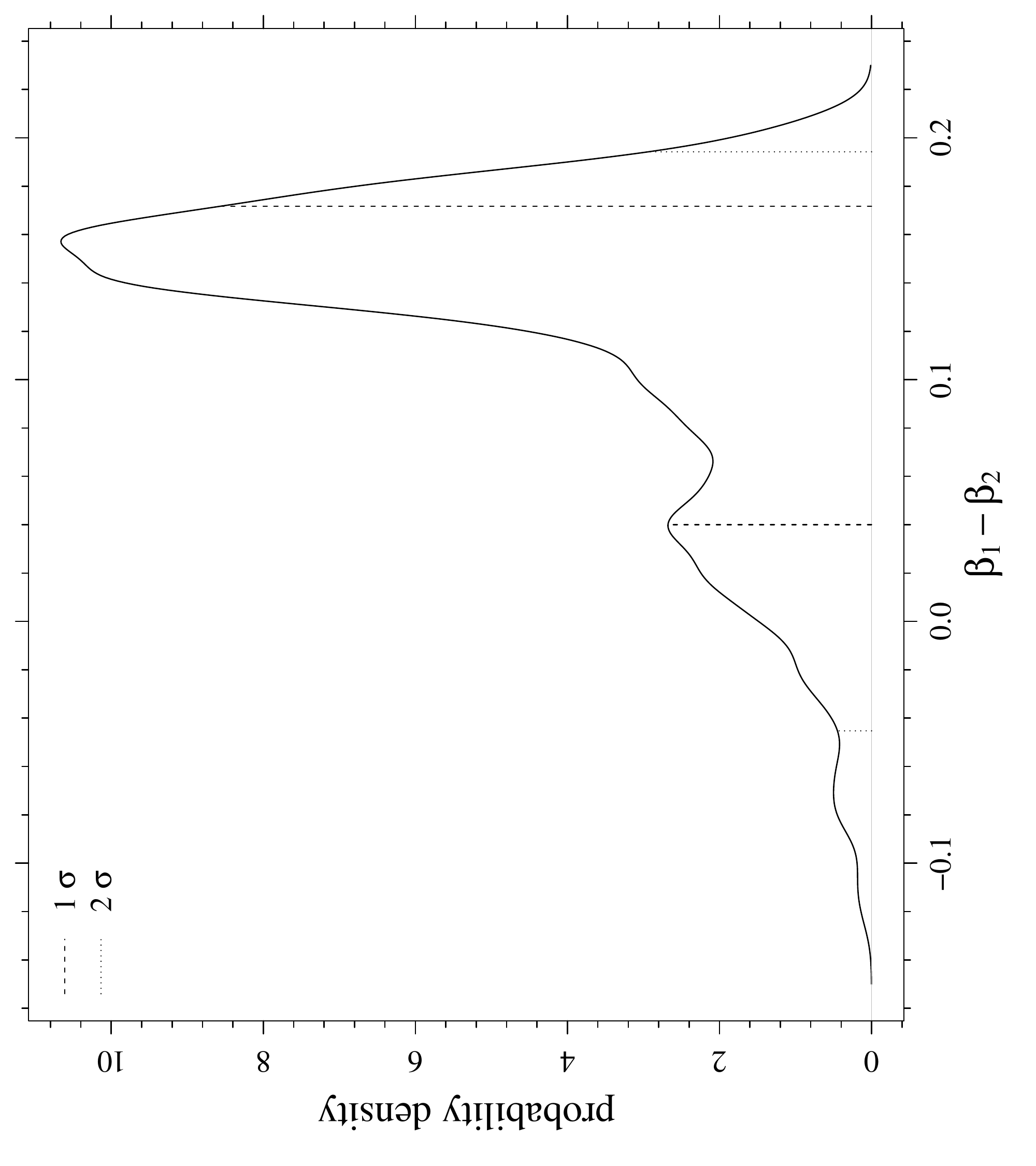}
        \includegraphics[height=6cm,angle=-90]{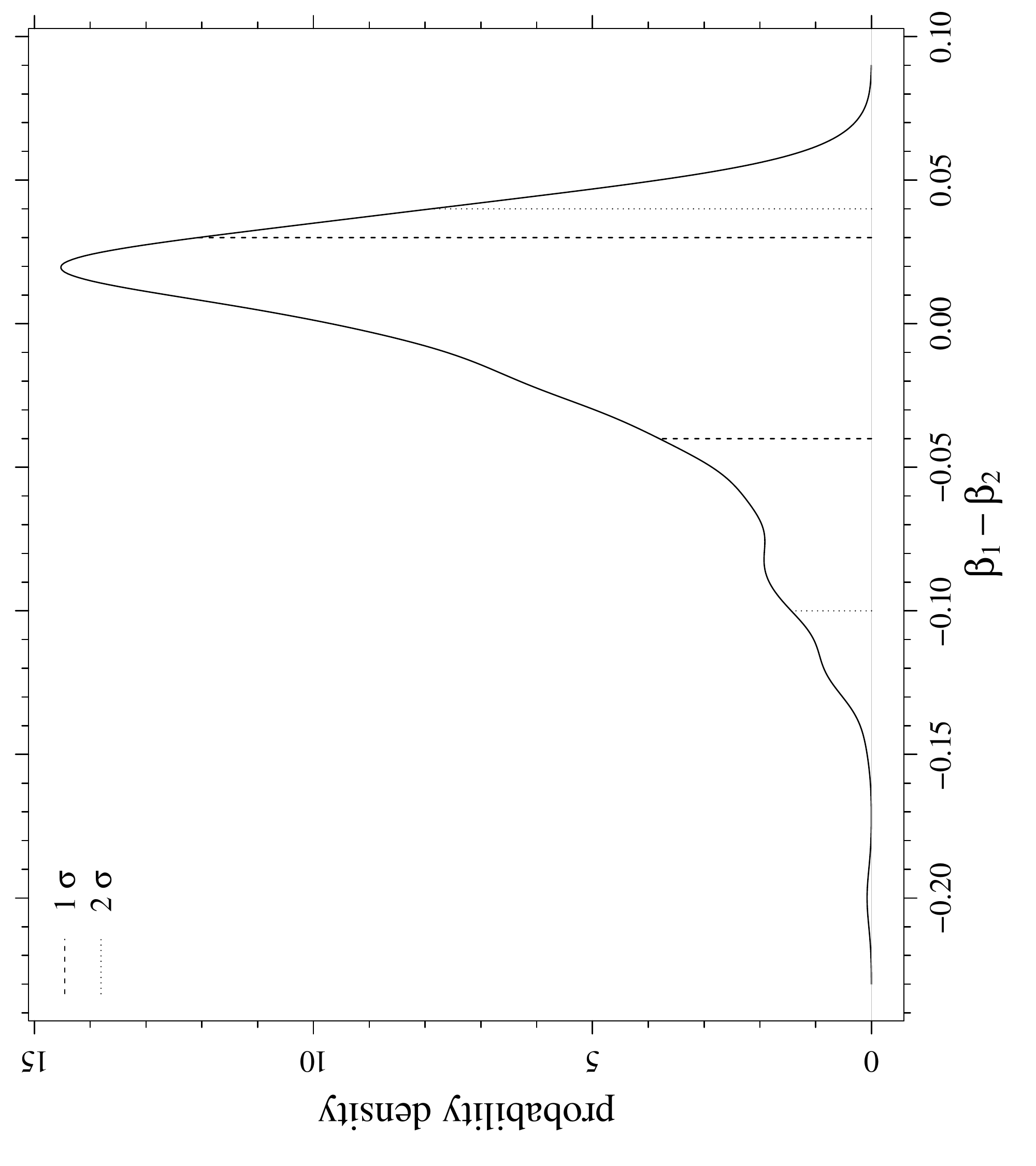}
        \includegraphics[height=6cm,angle=-90]{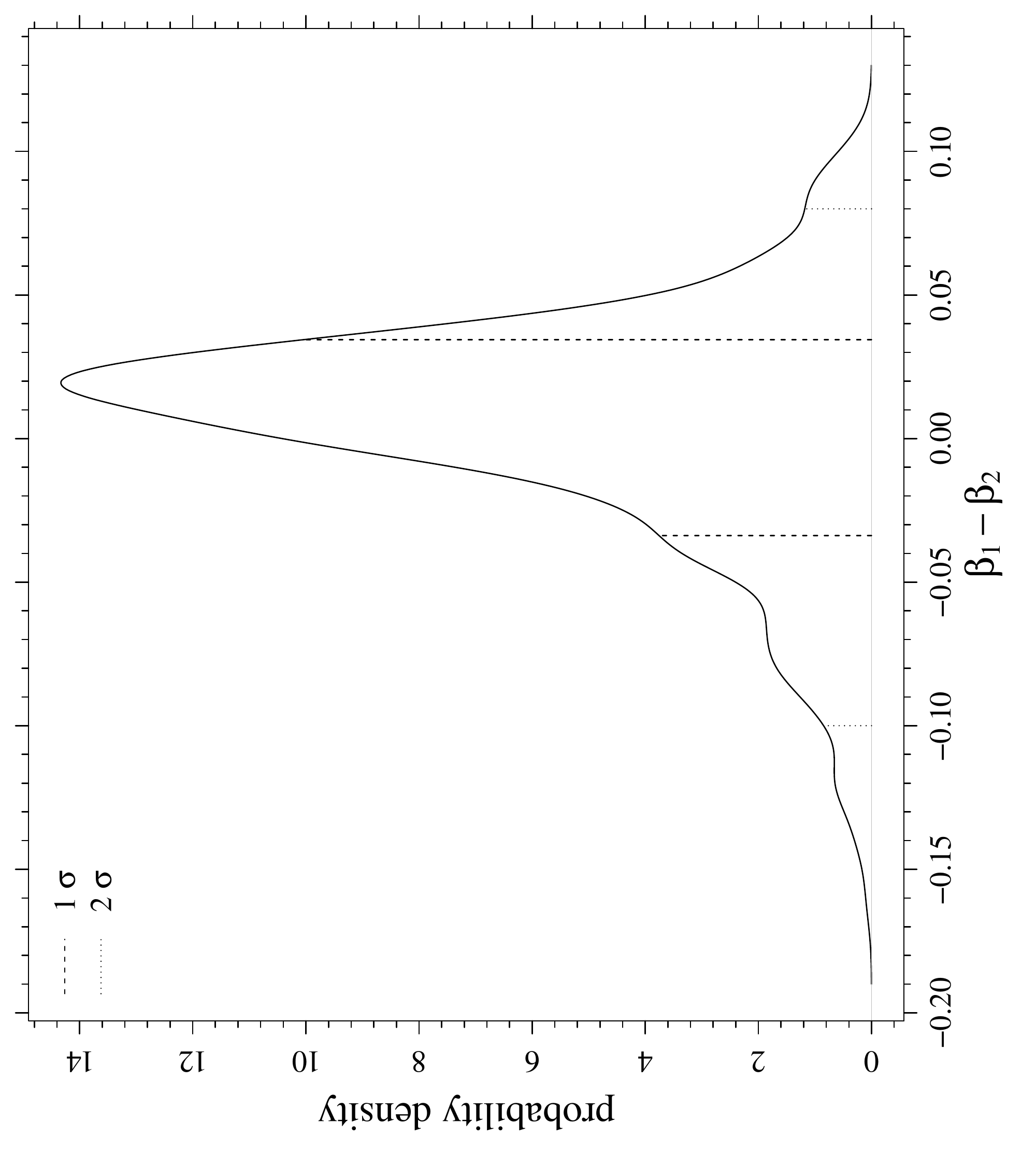}
        \caption{{\it Left}: Probability density of the difference of primary and secondary overshooting efficiencies under scenario $A$. {\it Middle}: As in the {\it left} panel, but for scenario $B$. {\it Right}: As in the {\it left} panel, but for scenario $C$. }
        \label{fig:OVdiff}
\end{figure*}

To describe the expected variability, we estimated the median value of the overshooting efficiencies $\beta_1$ and $\beta_2$ for primary and secondary stars for each synthetic system $N_1$. Table~\ref{tab:betasingle} lists the median $q_{50}$ of these $N_1$ values, as well as the 16th and 84th quantiles, $q_{16}$ and $q_{84}$ , respectively.
Figure~\ref{fig:OVdiff} shows the probability density of the difference $\beta_1-\beta_2$ for scenarios $A$, $B$, and $C$. The worst case occurs under scenario $A$, because the probability density of the difference peaks at 0.14 and shows very large variance. The large offset from zero occurs because the median $\beta_1$ is at 0.30, very far from the reference value 0.16. For systems like this it appears that it is just not possible to establish the two overshooting efficiencies individually.

In the other scenarios, the two individual overshooting efficiencies are less biased, and overall a $1 \sigma$ random variability of 0.03 is seen. In both cases, however, a long tail towards lower values (i.e. $\beta_2$ higher than $\beta_1$) is present: the 2.5th quantile (corresponding to a $2 \sigma$ variability) is found at $-0.10$.
The median individual overshooting efficiencies show a bias similar to those reported in Table.~\ref{tab:mainres}, with a tendency to even smaller overshooting efficiencies. Therefore -- while in the advanced evolutionary stages $B$ and $C$ it is possible to obtain an almost unbiased estimate of the differences between $\beta_1$ and $\beta_2$ -- individual $\beta$ estimates show a variance than can make it difficult to obtain a reliable calibration of the  overshooting efficiency with the stellar mass. A study based on a very large sample could in principle average out these random variations. However, even a large sample could not remedy the bias towards lower $\beta$ values.

\section{Conclusions}\label{sec:conclusions}

Following the theoretical analysis by \citet{overshooting} on the difficulties of obtaining a reliable calibration of the core overshooting efficiency for low-mass main sequence stars, the present work addresses the same question but for more massive and more evolved stars.
We focussed on a test case synthetic binary system composed of a 2.50 $M_{\sun}$ primary star coupled with a 2.38 $M_{\sun}$ secondary star. The selected case study is very similar in masses and metallicity to one of the brightest and most deeply studied binary systems in the sky, Capella ($\alpha$ Aur). We evaluated the theoretical foundations of calibrating the convective core overshooting efficiency from observational binary systems together with the system age determination. 

We put particular care into the statistical treatment of both the parameter estimates and their errors. To do this we used the SCEPtER pipeline \citep{scepter1,eta,binary}, a maximum likelihood procedure relying on a dense grid of stellar models, covering a wide range in both chemical abundances and in overshooting efficiencies. 
We selected three scenarios to address various evolutionary stages: the position of the primary is chosen at the end of the MS (scenario $A$), at the RGB start ($B$), or in the central helium burning ($C$).
For each of these scenarios, we adopted a two-stage Monte Carlo sampling procedure: a first step accounted for the random error in the observational data, which are otherwise supposed to be accurate. A second step was performed to recover the best fit stellar parameters from these perturbed systems. 

Adopting typical observational uncertainties of 100 K in the effective temperatures, 0.1 dex in [Fe/H], 1\% in the masses, and 0.5\% in the radii, we found that the recovered age and overshooting efficiency are biased towards low values in all three scenarios. The underestimation is particularly relevant for scenario $C$, reaching $-8.5\%$ in age and $-0.04$ in the overshooting parameter $\beta,$ that is, a relative error of about $-25\%$. In the other scenarios, an undervaluation of the age by about 4\% occurs. The origin of the bias is not linked to the particular technique of parameter recovery, but is due to the position of the stellar tracks in the hyperspace of the parameters adopted for the fitting, and is therefore unavoidable for the constraints considered here. The most important result, however, is the large variability in the fitted values owing to the observational errors. For  scenarios $A$ and $C$, the $\beta$ values showed extreme variations from one Monte Carlo simulation to another. This result strongly suggests that a calibration obtained from systems similar to the ones simulated in this work is ill-advised, since the results can be simply a product of random variation. 

To explore the impact of better observational constraints, we repeated the fitting by assuming a very small error of 0.1\% in the stellar masses (scenarios $A_M$, $B_M$, and $C_M$). As a result, the biases remain nearly unchanged but the variability is suppressed by a factor of about two.

We also explored the relevance of systematic mismatch between observed data and the recovery grid. We showed that the calibration from a binary system assuming an offset in the effective temperature of the stars of $\pm 150$ K from the grid leads to an impossibility in recovering the true system  values. 

Furthermore, the robustness of the distinction of different overshooting parameter values was also addressed by comparing the reconstruction of models sampled at $\beta = 0.0$ with that of models at $\beta = 0.16$. We obtained that the power of the distinction is generally high (greater than 85\%), even if potential problems can arise under scenario $C$, due to the presence of multiple solutions in the case of the sampling at $\beta = 0.0$.    

Moreover, we tested the possibility of independently fitting the overshooting efficiency of the two stars. This turned out to be impossible in scenario $A;$ due to a relaxation of the constraint of common $\beta$, the best fit overshooting parameter of the primary is strongly biased. In the other two scenarios, the fitting procedure recovers an almost identical median $\beta$ for the two stars, but with a non-negligible variability and a long tail towards the overestimation of the overshooting efficiency of the secondary star. These results advise caution when relying on individually fitted $\beta$ for stars in a binary system, as could be required for systems with stars of significantly different masses.

Most of the results presented in this paper were obtained assuming either perfectly accurate observations -- affected only by random perturbations -- or a perfect description of the binary system from the theoretical stellar tracks adopted in the recovery.  For real stars 
in a binary system these hypotheses are surely too optimistic. When the stated assumptions do not hold, it is apparent that the calibrated parameters can be potentially much more biased, as we showed for the effective temperature.  

In light of the present study and of those by \citet{binary} and \citet{overshooting}, it seems that the calibration procedure provides biased results, owing to the intertwined effects of all the unconstrained parameters, at least for systems affected by typical observational uncertainties. 
A minimum systematic uncertainty of about $\pm 0.03$ ($\pm 20\%$) should be considered on the recovered convective core overshooting parameter for stars evolved past the MS. For stars near the central hydrogen exhaustion it seems that the overshooting efficiency cannot be recovered in a reliable way at least for models studied in the present work and in \citet{overshooting}. 
The actual relevance of this uncertainty is obviously related to the purpose of the fit; while it impacts in a relevant way on every single system fit, nonetheless it allows a global approach, similar to those performed by \citet{Claret2016, Claret2017} in the examination of a possible statistical dependence of the convective core overshooting parameter on the mass of the stars. In these studies the proposed plateau for the $\beta$ value is about 0.20 for stars more massive than 2 $M_{\sun}$,  a  value we showed to be  distinguishable from no overshooting, apart from in peculiar cases. 

Further theoretical studies are needed to clarify whether the conclusions of this work are valid for binary systems in general or are simply relevant to those systems studied here.

\begin{acknowledgements}
We thank our anonymous referee for the many useful suggestions that helped us to improve the manuscript.
This work has been supported by PRA Universit\`{a} di Pisa 2016 
(\emph{Stelle di piccola massa: le pietre miliari dell'archeologia galattica}, PI: S. Degl'Innocenti) and by INFN (\emph{Iniziativa specifica TAsP}).
\end{acknowledgements}

\bibliographystyle{aa}
\bibliography{biblio}

\begin{appendix}

\section{Random-effect models}\label{app:raneff}        

As described in Sect.~\ref{sec:method}, a two-stage approach was adopted in the Monte Carlo simulations. First, $N_1$ artificial systems were generated and subjected to perturbation to account for the observational errors. Second, all these systems were fitted by the algorithm described in Sect.~\ref{sec:fittingML}, which adopts $N_2$ perturbed replicates of each system to evaluate the statistical errors of the estimated parameters.
        
Let us denote by $i$ the artificial system ($i$ = 1, $\ldots$, $N_1$), and by $j$ the replicate inside a system ($j$ = 1, $\ldots$, $N_2$). Without a loss of generality, let us consider the age
of the system as the dependent variable $Y$ .
A fixed-effect model for the age with respect to the artificial system classification would be specified as \citep[see e.g.][]{snedecor1989, simar}:
\begin{equation}
Y_{ij} = \mu + \alpha_i + \varepsilon_{ij} , 
\end{equation}
where $\mu$ is the overall mean of the data, $\alpha_i$ are the parameters (to be estimated from the model) for the difference in age among artificial systems, and  $\varepsilon_{ij} \sim N(0, \sigma^2)$ is the Gaussian error term.
This model corresponds to a classical analysis of variance (ANOVA), and is suitable if the individual differences in the age among the exact artificially selected systems are of interest. 

However, these systems are only a random sample of the possible ones than can be generated by the Monte Carlo procedure. Hence, it is more relevant to estimate the {\it variability} in the mean age due to the random sampling. 
This is achieved by adopting a random effects model:
\begin{equation}
Y_{ij} = \mu + A_i + \varepsilon_{ij}\label{eq:raneff} ,
\end{equation}
where $\varepsilon_{ij} \sim N(0, \sigma^2)$ and, differing from the previous model, $A_i \sim N(0, \sigma_g^2)$ are random variables.
The estimate of $\sigma$ and $\sigma_g$ are the outcome of the model fitting.

The fit of the model in Eq.~(\ref{eq:raneff}) was performed using a restricted maximum likelihood technique adopting the library {\it lme4} in R 3.3.1 \citep{lme4, R}.
Further details on the method and on its theoretical assumptions can be found in \citet{Laird1982, venables2002modern, lme4}.
        
\end{appendix}

\end{document}